\documentclass[twocolumn]{aastex631} 

\usepackage{lipsum} 
\usepackage{tabularx}
\usepackage{amsmath}
\usepackage{threeparttable}
\usepackage{cleveref}
\usepackage{tablefootnote} 

\begin{document}

\title{
Tracing the Cosmic Evolution of the Cool Circumgalactic Medium of Luminous Red Galaxies with DESI Year 1 Data
}


\correspondingauthor{Yu-Ling Chang}
\email{ecylchang@ntu.edu.tw}

\author[0000-0002-0196-3496]{Yu-Ling Chang}
\affiliation{Graduate Institute of Astrophysics, National Taiwan University, No. 1, Sec. 4, Roosevelt Road, Taipei 10617, Taiwan}

\author[0000-0001-8857-7020]{Ting-Wen Lan}
\affiliation{Graduate Institute of Astrophysics, National Taiwan University, No. 1, Sec. 4, Roosevelt Road, Taipei 10617, Taiwan}
\affiliation{Department of Physics, National Taiwan University, No. 1, Sec. 4, Roosevelt Rd., Taipei 10617, Taiwan}
\affiliation{Institute of Astronomy and Astrophysics, Academia Sinica, No. 1, Sec. 4, Roosevelt Rd., Taipei 10617, Taiwan}

\author[0000-0002-7738-6875]{J. Xavier Prochaska}
\affiliation{University of California, Santa Cruz, 1156 High Street, Santa Cruz, CA 95064, USA}
\affiliation{Kavli Institute for the Physics and Mathematics of the Universe, 5-1-5 Kashiwanoha, Kashiwa, 277-8583, Japan}

\author[0000-0002-2949-2155]{Malgorzata Siudek}
\affiliation{Institute of Space Sciences, ICE-CSIC, Campus UAB, Carrer de Can Magrans s/n, 08913 Bellaterra, Barcelona, Spain}
\affiliation{Instituto Astrofisica de Canarias, Av. Via Lactea s/n, 38205 La Laguna, Spain}

\author{J.~Aguilar}
\affiliation{Lawrence Berkeley National Laboratory, 1 Cyclotron Road, Berkeley, CA 94720, USA}

\author[0000-0001-6098-7247]{S.~Ahlen}
\affiliation{Department of Physics, Boston University, 590 Commonwealth Avenue, Boston, MA 02215 USA}

\author[0000-0003-2923-1585]{Abhijeet Anand}
\affiliation{Lawrence Berkeley National Laboratory, 1 Cyclotron Road, Berkeley, CA 94720, USA}

\author[0000-0001-9712-0006]{D.~Bianchi}
\affiliation{Dipartimento di Fisica ``Aldo Pontremoli'', Universit\`a degli Studi di Milano, Via Celoria 16, I-20133 Milano, Italy}
\affiliation{INAF-Osservatorio Astronomico di Brera, Via Brera 28, 20122 Milano, Italy}

\author{D.~Brooks}
\affiliation{Department of Physics \& Astronomy, University College London, Gower Street, London, WC1E 6BT, UK}

\author[0000-0001-7316-4573]{F.~J.~Castander}
\affiliation{Institut d'Estudis Espacials de Catalunya (IEEC), c/ Esteve Terradas 1, Edifici RDIT, Campus PMT-UPC, 08860 Castelldefels, Spain}
\affiliation{Institute of Space Sciences, ICE-CSIC, Campus UAB, Carrer de Can Magrans s/n, 08913 Bellaterra, Barcelona, Spain}

\author{T.~Claybaugh}
\affiliation{Lawrence Berkeley National Laboratory, 1 Cyclotron Road, Berkeley, CA 94720, USA}

\author[0000-0002-1769-1640]{A.~de la Macorra}
\affiliation{Instituto de F\'{\i}sica, Universidad Nacional Aut\'{o}noma de M\'{e}xico,  Circuito de la Investigaci\'{o}n Cient\'{\i}fica, Ciudad Universitaria, Cd. de M\'{e}xico  C.~P.~04510,  M\'{e}xico}

\author{P.~Doel}
\affiliation{Department of Physics \& Astronomy, University College London, Gower Street, London, WC1E 6BT, UK}

\author[0000-0003-4992-7854]{S.~Ferraro}
\affiliation{Lawrence Berkeley National Laboratory, 1 Cyclotron Road, Berkeley, CA 94720, USA}
\affiliation{University of California, Berkeley, 110 Sproul Hall \#5800 Berkeley, CA 94720, USA}

\author[0000-0002-3033-7312]{A.~Font-Ribera}
\affiliation{Institut de F\'{i}sica d’Altes Energies (IFAE), The Barcelona Institute of Science and Technology, Edifici Cn, Campus UAB, 08193, Bellaterra (Barcelona), Spain}

\author[0000-0002-2890-3725]{J.~E.~Forero-Romero}
\affiliation{Departamento de F\'isica, Universidad de los Andes, Cra. 1 No. 18A-10, Edificio Ip, CP 111711, Bogot\'a, Colombia}
\affiliation{Observatorio Astron\'omico, Universidad de los Andes, Cra. 1 No. 18A-10, Edificio H, CP 111711 Bogot\'a, Colombia}

\author[0000-0001-9632-0815]{E.~Gaztañaga}
\affiliation{Institut d'Estudis Espacials de Catalunya (IEEC), c/ Esteve Terradas 1, Edifici RDIT, Campus PMT-UPC, 08860 Castelldefels, Spain}
\affiliation{Institute of Cosmology and Gravitation, University of Portsmouth, Dennis Sciama Building, Portsmouth, PO1 3FX, UK}
\affiliation{Institute of Space Sciences, ICE-CSIC, Campus UAB, Carrer de Can Magrans s/n, 08913 Bellaterra, Barcelona, Spain}

\author[0000-0003-3142-233X]{S.~Gontcho A Gontcho}
\affiliation{Lawrence Berkeley National Laboratory, 1 Cyclotron Road, Berkeley, CA 94720, USA}
\affiliation{University of Virginia, Department of Astronomy, Charlottesville, VA 22904, USA}

\author{G.~Gutierrez}
\affiliation{Fermi National Accelerator Laboratory, PO Box 500, Batavia, IL 60510, USA}

\author[0000-0001-9822-6793]{J.~Guy}
\affiliation{Lawrence Berkeley National Laboratory, 1 Cyclotron Road, Berkeley, CA 94720, USA}

\author[0000-0002-6550-2023]{K.~Honscheid}
\affiliation{Center for Cosmology and AstroParticle Physics, The Ohio State University, 191 West Woodruff Avenue, Columbus, OH 43210, USA}
\affiliation{Department of Physics, The Ohio State University, 191 West Woodruff Avenue, Columbus, OH 43210, USA}
\affiliation{The Ohio State University, Columbus, 43210 OH, USA}

\author[0000-0003-0201-5241]{R.~Joyce}
\affiliation{NSF NOIRLab, 950 N. Cherry Ave., Tucson, AZ 85719, USA}

\author[0000-0002-0000-2394]{S.~Juneau}
\affiliation{NSF NOIRLab, 950 N. Cherry Ave., Tucson, AZ 85719, USA}

\author[0000-0001-6356-7424]{A.~Kremin}
\affiliation{Lawrence Berkeley National Laboratory, 1 Cyclotron Road, Berkeley, CA 94720, USA}

\author{O.~Lahav}
\affiliation{Department of Physics \& Astronomy, University College London, Gower Street, London, WC1E 6BT, UK}

\author[0000-0002-6731-9329]{C.~Lamman}
\affiliation{The Ohio State University, Columbus, 43210 OH, USA}

\author[0000-0003-1838-8528]{M.~Landriau}
\affiliation{Lawrence Berkeley National Laboratory, 1 Cyclotron Road, Berkeley, CA 94720, USA}

\author[0000-0001-7178-8868]{L.~Le~Guillou}
\affiliation{Sorbonne Universit\'{e}, CNRS/IN2P3, Laboratoire de Physique Nucl\'{e}aire et de Hautes Energies (LPNHE), FR-75005 Paris, France}

\author[0000-0003-1887-1018]{M.~E.~Levi}
\affiliation{Lawrence Berkeley National Laboratory, 1 Cyclotron Road, Berkeley, CA 94720, USA}

\author[0000-0003-4962-8934]{M.~Manera}
\affiliation{Departament de F\'{i}sica, Serra H\'{u}nter, Universitat Aut\`{o}noma de Barcelona, 08193 Bellaterra (Barcelona), Spain}
\affiliation{Institut de F\'{i}sica d’Altes Energies (IFAE), The Barcelona Institute of Science and Technology, Edifici Cn, Campus UAB, 08193, Bellaterra (Barcelona), Spain}

\author[0000-0002-1125-7384]{A.~Meisner}
\affiliation{NSF NOIRLab, 950 N. Cherry Ave., Tucson, AZ 85719, USA}

\author{R.~Miquel}
\affiliation{Instituci\'{o} Catalana de Recerca i Estudis Avan\c{c}ats, Passeig de Llu\'{\i}s Companys, 23, 08010 Barcelona, Spain}
\affiliation{Institut de F\'{i}sica d’Altes Energies (IFAE), The Barcelona Institute of Science and Technology, Edifici Cn, Campus UAB, 08193, Bellaterra (Barcelona), Spain}

\author[0000-0001-9070-3102]{S.~Nadathur}
\affiliation{Institute of Cosmology and Gravitation, University of Portsmouth, Dennis Sciama Building, Portsmouth, PO1 3FX, UK}

\author[0000-0001-8684-2222]{J.~ A.~Newman}
\affiliation{Department of Physics \& Astronomy and Pittsburgh Particle Physics, Astrophysics, and Cosmology Center (PITT PACC), University of Pittsburgh, 3941 O'Hara Street, Pittsburgh, PA 15260, USA}

\author[0000-0002-0644-5727]{W.~J.~Percival}
\affiliation{Department of Physics and Astronomy, University of Waterloo, 200 University Ave W, Waterloo, ON N2L 3G1, Canada}
\affiliation{Perimeter Institute for Theoretical Physics, 31 Caroline St. North, Waterloo, ON N2L 2Y5, Canada}
\affiliation{Waterloo Centre for Astrophysics, University of Waterloo, 200 University Ave W, Waterloo, ON N2L 3G1, Canada}

\author{C.~Poppett}
\affiliation{Lawrence Berkeley National Laboratory, 1 Cyclotron Road, Berkeley, CA 94720, USA}
\affiliation{Space Sciences Laboratory, University of California, Berkeley, 7 Gauss Way, Berkeley, CA  94720, USA}
\affiliation{University of California, Berkeley, 110 Sproul Hall \#5800 Berkeley, CA 94720, USA}

\author[0000-0001-7145-8674]{F.~Prada}
\affiliation{Instituto de Astrof\'{i}sica de Andaluc\'{i}a (CSIC), Glorieta de la Astronom\'{i}a, s/n, E-18008 Granada, Spain}

\author[0000-0001-6979-0125]{I.~P\'erez-R\`afols}
\affiliation{Departament de F\'isica, EEBE, Universitat Polit\`ecnica de Catalunya, c/Eduard Maristany 10, 08930 Barcelona, Spain}

\author{G.~Rossi}
\affiliation{Department of Physics and Astronomy, Sejong University, 209 Neungdong-ro, Gwangjin-gu, Seoul 05006, Republic of Korea}

\author[0000-0002-9646-8198]{E.~Sanchez}
\affiliation{CIEMAT, Avenida Complutense 40, E-28040 Madrid, Spain}

\author{D.~Schlegel}
\affiliation{Lawrence Berkeley National Laboratory, 1 Cyclotron Road, Berkeley, CA 94720, USA}

\author{M.~Schubnell}
\affiliation{Department of Physics, University of Michigan, 450 Church Street, Ann Arbor, MI 48109, USA}
\affiliation{University of Michigan, 500 S. State Street, Ann Arbor, MI 48109, USA}

\author{D.~Sprayberry}
\affiliation{NSF NOIRLab, 950 N. Cherry Ave., Tucson, AZ 85719, USA}

\author[0000-0003-1704-0781]{G.~Tarl\'{e}}
\affiliation{University of Michigan, 500 S. State Street, Ann Arbor, MI 48109, USA}

\author{B.~A.~Weaver}
\affiliation{NSF NOIRLab, 950 N. Cherry Ave., Tucson, AZ 85719, USA}

\author[0000-0001-5381-4372]{R.~Zhou}
\affiliation{Lawrence Berkeley National Laboratory, 1 Cyclotron Road, Berkeley, CA 94720, USA}

\author[0000-0002-6684-3997]{H.~Zou}
\affiliation{National Astronomical Observatories, Chinese Academy of Sciences, A20 Datun Rd., Chaoyang District, Beijing, 100012, P.R. China}

\begin{abstract}
We investigate the properties of the cool circumgalactic medium (CGM) of massive galaxies and their cosmic evolution. By using the year 1 dataset of luminous red galaxies (LRGs) and QSOs from the Dark Energy Spectroscopic Instrument survey, we construct a sample of approximately 800,000 galaxy-quasar pairs and measure the radial distribution and kinematics of the cool gas traced by Mg~{\small II} absorption lines as a function of galaxy properties from redshift 0.4 to redshift 1.2. 
Our results show that the covering fraction of the cool gas around LRGs increases with redshift, following a trend similar to the global evolution of galaxy star formation rate.
At small radii ($\lesssim 0.3~r_{\rm vir}$), the covering fraction anti-correlates with stellar mass, suggesting that mass-dependent processes 
suppress the cool gas content in the inner region. 
In addition, we measure the gas dispersion by modeling the velocity distribution of absorbers with a narrow and a broad components -- $\rm \sigma_{narrow}\sim160 \,  km/s$ and $\rm \sigma_{broad}\sim380\, km/s$ --
and quantify their relative contributions. 
The results show that the broad component becomes more prominent in the outer region, and its relative importance in the central region grows with increasing stellar mass. 
Finally, we discuss possible origins of the cool gas around massive galaxies, including the contribution of satellite galaxies and the precipitation scenario. 

\end{abstract}

\keywords{Spectroscopy(1558) --- Extragalactic astronomy(506) --- Circumgalactic medium(1879)}

\section{Introduction} \label{intro}

The circumgalactic medium \citep[CGM, see][for a review]{Tumlinson2017, Peroux2020, Faucher-Giguere2023} provides an essential gas-based perspective on how galaxies evolve. 
As the CGM harbors gas flows regulating mass and energy exchange between galaxies and their surroundings, the CGM properties offer crucial insights into the physical mechanisms shaping galaxy growth \citep[e.g.,][]{Weiner2009, Tumlinson2011, Rubin2012, 
Werk2014, Peeples2014}. Absorption line spectroscopy has been widely used to probe the CGM properties \citep[e.g.,][]{Bergeron1986, Steidel1994, Williams2005, Werk2013, Chen2020, Tchernyshyov2023, Zheng2024, Ng25, Lan2025}, as the diffuse gas is difficult to detect in emission \citep[e.g.,][]{Corlies2016, Dutta2024a, Nielsen2024}. With background sources intercepting the CGM of foreground galaxies, one can detect the absorption features imprinted in the spectra of the background sources, infer the gas properties and investigate the connections between galaxies and their gas. 


Among the absorption line transitions, the Mg~{\small II}~$\lambda\lambda~2796, 2803$ doublet
which traces cool gas with $T \sim 10^4 \, {\rm K}$, is one of the strongest absorption features which can be detected in optical wavelengths at moderate redshifts $(0.3 < z < 2.5)$. 
Therefore, it has frequently
been used to trace the cool CGM over cosmic time \citep[e.g.,][]{Bergeron1986, Churchill1999, Churchill2000, Churchill2001, Churchill2003, Nestor2005, Nielsen2013a, Zhu2013b, Raghunathan2016, Chen2017, Zou2021, Napolitano2024, Bouche2025}. 
Furthermore,  previous studies have 
investigated the cool CGM properties as a function of galaxy properties, including stellar mass, star-formation rate, impact parameter, azimuthal angle, morphology, AGN activity and redshifts \citep[e.g.,][]{Chen2010, Nielsen2013b, Lan14, Raghunathan2016, Nielsen2016, Chen2017, Kauffmann2017, Rubin2018, Lan2018, Anand2021, Anand2022, Zou2024, Chang2024, Chen2025}. 
One puzzling result has emerged from these studies --- a non-negligible amount of cool gas is detected around massive passive galaxies \citep[e.g.,][]{Gauthier2011, Zhu2014, Lan14, Huang2016, Chen2018, Lan2020, Huang2021, Dutta2020}, contrary to the expectation that 
their quenched star formation implies little cool gas if any \citep[e.g.,][]{Saintonge2017, Catinella2018}. 
Several studies have focused on exploring the cool gas around massive passive galaxies.
For example, \citet{Huang2016} studied the cool CGM properties for quiescent galaxies using $\sim 38,000$ luminous red galaxy (LRG) - QSO pairs from Sloan Digital Sky Survey (SDSS) DR12 \citep{SDSSDR12} and found suppressed Mg~{\small II} velocity dispersions and elevated covering fractions along the major axis of [O~{\small II}]-emitting LRGs, suggesting contributions from thermal instabilities and satellite accretion along filaments. 
\citet{Chen2018} and \citet{Zahedy2019} combined {\it Hubble Space Telescope} Cosmic Origins Spectrograph \citep[COS,][]{Green2012} data and the MIKE and HIRES echelle \citep{Vogt1994, Bernstein2003} spectra\footnote{The MIKE and HIRES spectrographs are installed on the Magellan Clay telescope and the Keck I telescope, respectively.} to study the physical conditions and element abundance of the gas surrounding 16 COS LRGs. 
Their findings indicate a chemically enriched, multiphase CGM consistent with precipitation cooling frameworks \citep{Werner2014, Voit2015a}. 
On the other hand, \citet{Lan2020} correlated Mg~{\small II} absorbers from \citet{Zhu2013b} with galaxies in the DESI Legacy Survey \citep{Dey2019}, probing the properties of these absorbers in relation to various galaxy properties. 
Their results indicate that the cool gas in star-forming satellites is sufficient to explain a large fraction of the cool gas around passive galaxies. 
Complementary results were reported by \citet{Smailagic2023}, who found that most LRG–CGM systems show weak or undetectable O~{\small VI} absorption, with the few strong cases likely arising from star-forming neighbors in LRG groups. 
Recently, \citet{Staffehl25} traced the origin of cool gas in present-day clusters using the TNG-Cluster simulation \citep{Nelson2024}. 
From their simulation, most cool gas originates from pre-cooled material carried by infalling satellites and other halos, with a secondary contribution from in-situ cooling of previously accreted hot gas.  



While these studies have explored various galaxy properties, one key observed property is still rarely investigated, i.e. the cosmic evolution of the cool gas around massive galaxies, which can offer further insights on the origins of the cool gas and how the cool gas co-evolves with those massive galaxies through cosmic time. 
Satellite-based scenarios, for example, predict a redshift evolution of the cool-gas covering fraction, as shown in cosmological simulations \citep{Hafen20, Nelson2020}. 
Large spectroscopic surveys, such as the Dark Energy Spectroscopic Instrument (DESI) survey \citep{DESI2016a, DESI_instr, DESI_DR1}, now offer the required large statistical samples across a wide range of redshift and enable studies of probing the redshift evolution of galaxies and CGM connections. 



In this work, we perform the first systematic investigation of the cool CGM properties of DESI LRGs across cosmic time from redshift $0.4-1.2$. Utilizing the largest spectroscopic dataset from DESI, we construct a sample of $\sim 800,000$ LRG-QSO pairs and explore (1) the correlations between galaxy and gas properties and (2) the redshift evolution of these correlations, revealing trends that provide constraints on the origins of the cool gas and the mechanisms regulating their properties. 



This paper is organized in the following way: 
we describes our dataset in Section~\ref{data}, and present the construction of galaxy-quasar pairs and detection of Mg~{\small II} absorbers in Section~\ref{analysis}. 
We show our results in Section~\ref{results} and discuss the implications in Section~\ref{discussion}. 
We conclude this work in Section~\ref{conclusion}. 
Throughout the paper, we adopt a flat-$\Lambda$CDM cosmology with $\Omega_{\rm m}=0.3~{\rm and}~H_{0}=70~{\rm km}~{\rm s}^{-1}~{\rm Mpc}^{-1}$. 

\begin{figure*}
\begin{center}
\includegraphics[width=1.\linewidth]{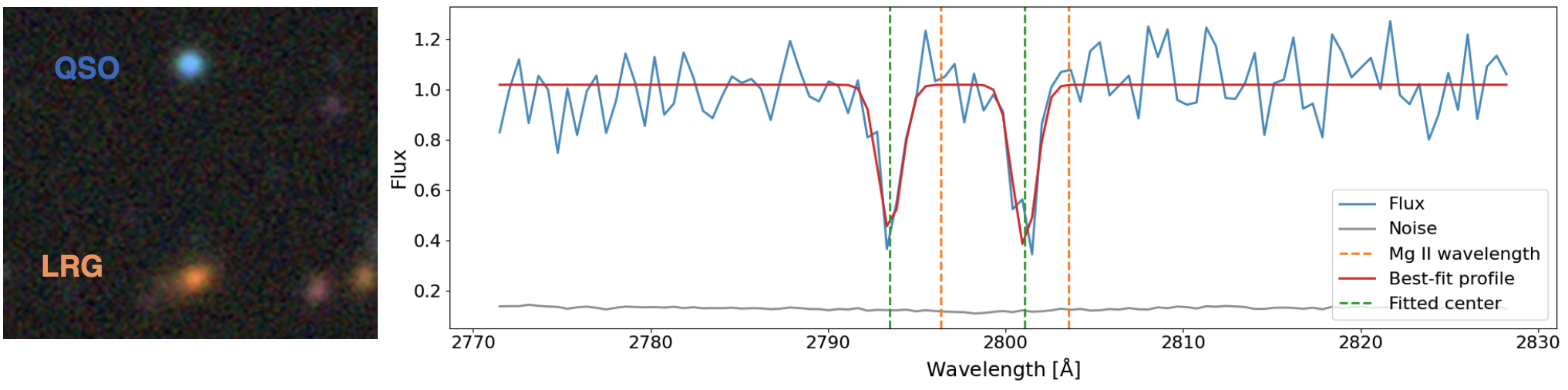}
\end{center}
\caption{Example of a LRG–QSO pair with detected Mg~{\small II} absorption. Left: Image of the LRG–QSO pair from the Legacy Survey DR9 \citep{Dey2019}. Right: The corresponding DESI spectrum of the background QSO in the LRG rest frame, showing the Mg II absorption line. The orange dashed lines mark the wavelengths of the Mg~{\small II} doublet, while the green dashed lines mark the centers of the fitted absorption lines. This difference indicates the gas velocity relative to the galaxy.}
\label{pair_fitting_example}
\end{figure*}

\section{Data} \label{data}
We make use of data collected by the Dark Energy Spectroscopic Instrument (DESI) on the Mayall 4-meter telescope at Kitt Peak National Observatory \citep{Levi2013}. 
DESI is designed to study the nature of dark energy by measuring the expansion history of the universe with tens of millions of spectra of extragalactic sources \citep{DESI2016a, DESI2016b}.
To observe a large number of sources simultaneously, DESI is equipped with a highly multiplexed spectrograph featuring a unique focal plane system with 5,000 fibers and robotic positioners \citep{DESI_focul_plane, Miller2024, Poppett2024}. 
DESI spectra were observed from $3600 - 9800{\rm \AA}$, with a spectral resolution of 2000-5000 and a linear sampling of $0.8\rm \, \AA$ \citep{DESI_instr,DESI_pipeline}. 
The Redrock pipeline\footnote{\url{https://github.com/desihub/redrock}} \citep[][and Bailey et al. in prep]{Brodzeller2023, Anand2024} then automatically classifies each object and measures its spectroscopic redshift. 

To optimize data collection, the DESI survey utilizes both bright and dark time to observe different types of sources \citep{DESI_survy_ops}, pre-selected from the DESI Legacy Imaging Surveys \citep{Dey2019}. 
The bright-time program comprises surveys of nearby bright sources, such as the Milky Way survey \citep{DESI_MWS} and the bright galaxy survey \citep[BGS,][]{DESI_BGS,DESI_BGS_AGN}. 
In contrast, fainter extragalactic sources, like luminous red galaxies \citep[LRGs,][]{DESI_LRG}, emission line galaxies \citep[ELGs,][]{DESI_ELG}, and quasars \citep{DESI_QSO}, are observed in the dark-time program. 
Before launching its five-year main survey in 2021, a survey validation (SV) stage was conducted to test target selection approaches \citep{DESI_target}, validate redshift performance, and optimize exposure times for each DESI target class \citep{DESI_SV, DESI_galaxy_VI, DESI_quasar_VI}.

\begin{figure*}
\begin{center}
\includegraphics[width=1.\linewidth]{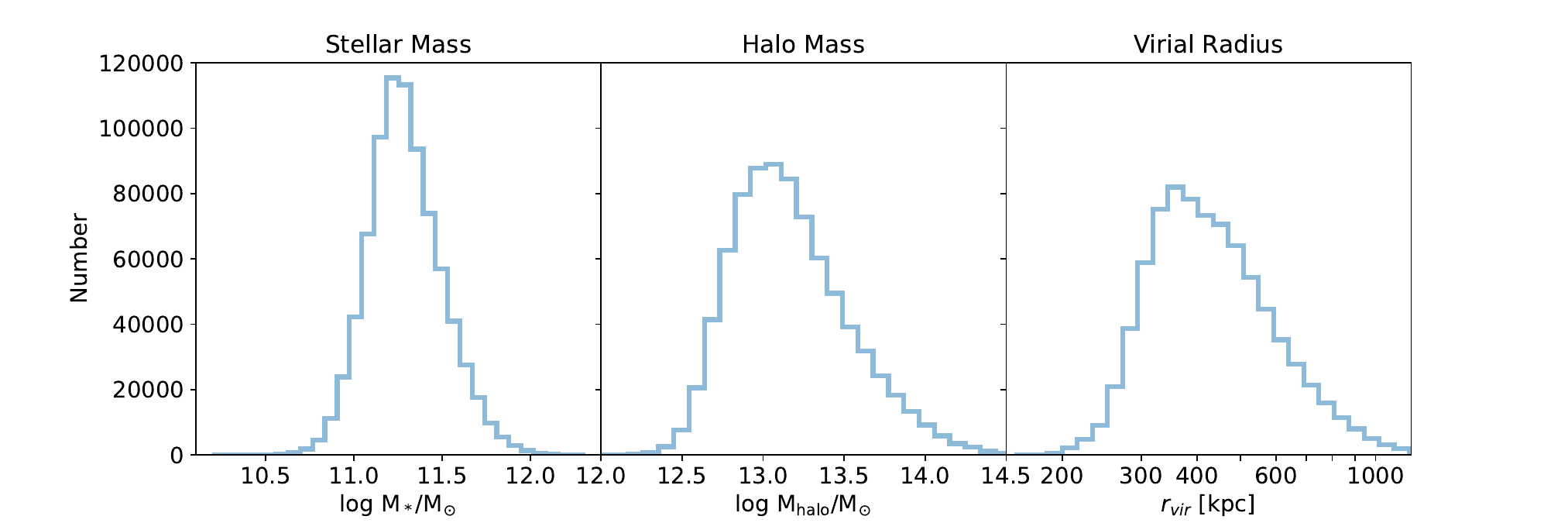}
\end{center}
\caption{Distributions of stellar masses (left panel), halo masses (middle panel) and virial radius (right panel). The stellar masses are derived from \texttt{CIGALE}, while the halo masses are derived from SHMR and HMF. See the text for more details about the estimation of halo masses and virial radius. }
\label{stellar_dist}
\end{figure*}

Here we use the LRG and quasar data \citep{Ross2025} from the Year 1 of the DESI main survey \citep{DESI_DR1}, as well as the SV \citep[Early Data Release,][]{DESI_EDR_cat}. 
The cosmological constraints from baryon acoustic oscillation measurements based on the first-year and three-year main survey data are presented in \citet{DESI2024VII, DESI_DR2_I, DESI_DR2_II}. 

\textbf{The DESI LRG survey:}
To primarily observe massive and passive galaxies from $z\sim0.3$ to $z\sim1.1$, DESI LRG targets are selected based on $r-{\rm W1}$ color and brightness. 
The typical redshift error for DESI LRGs is $\sim 50~{\rm km/s}$ with a catastrophic rate of $\leq 0.3\%$ \citep{DESI_DR1, DESI2025}. 
We use DESI LRGs with {\tt ZWARN} = 0\footnote{\url{https://desidatamodel.readthedocs.io/en/latest/bitmasks.html\#zwarn-bitmask}} and {\tt DELTACHI2} ($\Delta \chi ^2$)\footnote{$\chi^2$ difference between the second and first best-fit model} $> 15$, i.e.\ sources with robust redshift estimation \citep{DESI_LRG, DESI_galaxy_VI}. 
A redshift cut of $0.4 < z < 1.2$ is adopted, motivated by the DESI spectral coverage of Mg~{\small II} at the shortest wavelengths and 
to avoid the missed classification of LRGs at higher redshifts \citep{DESI_LRG}. 
The above selection leads to a sample of $\sim2.2\times10^{6}$ LRGs. 

\textbf{The DESI QSO survey:}
DESI selects QSO targets using both the random forest approach and infrared and optical colors \citep{DESI_QSO}. 
The redshift precision of DESI QSOs ranges from $\sim 20$ to $\sim 125~{\rm km\,s^{-1}}$ over $z \sim 0.8$–1.8 and improves at higher redshift, with catastrophic offsets ($>3000~{\rm km/s}$) at 0.7\% for $z<2.1$ and 1.8\% for $z>2.1$ \citep{DESI2025}.
We apply the same criteria {\tt ZWARN} = 0 and {\tt DELTACHI2} $> 15$ as for LRGs to ensure select quasars with robust redshift measurement \citep{DESI_galaxy_VI, DESI_quasar_VI}. 
We also remove quasars with spectra having a median signal-to-noise ratio (SNR) lower than 1 across the full wavelength range. 
The final quasar sample consists of approximately $1.2 \times 10^6$ sources. 

\section{Analysis} \label{analysis}
\subsection{Detection of Mg~{\small II} absorbers} \label{build_sample}
Our goal is to investigate the properties of the cool gas, traced by Mg~{\small II} absorption lines, associated with massive galaxies. 
Following \citet{Chang2024}, we search for Mg~{\small II} absorbers and measure their radial distribution and kinematics around DESI LRGs.

\subsubsection{Constructing a sample of LRG-QSO pairs} \label{selecting_pair}
First, we build a sample of LRGs with background quasars within a physical impact parameter of $1000$~kpc or a halo-scaled impact parameter of $2.5\,r_{vir}$, resulting in $1,712,316$ LRG-QSO pairs. 
The impact parameters ($r_p$) are the projected distances between the galaxies and the quasar sightlines. 
We further exclude pairs with potential Mg~{\small II} absorbers contaminated by (1) Mg~{\small II}-associated absorbers intrinsic to the background QSOs, (2) C~{\small III} emission lines (3) C~{\small IV} emission lines and blended absorption features within the Ly$\alpha$ forest \citep[e.g.,][]{Zhu2013b}. 
Specifically, we apply three velocity criteria to mask spectral regions with potential contamination from absorptions close to Mg\,{\small II}, C\,{\small III}, or C\,{\small IV} emission lines: $\delta v \le 6000$, $|\delta v| \le 8000$, and $|\delta v| \le 10\,000~{\rm km\,s^{-1}}$, respectively.
Here $\delta v = c\Delta z/(1+z_{\rm QSO})$ is the LoS velocity separation between the QSO redshift and the relevant critical redshift, with $\Delta z = z_{\rm QSO}-z_{\rm crit}$. The critical redshift, $z_{\rm crit}$, is defined as the redshift at which a QSO-frame transition would coincide with the expected Mg~{\small II} absorption wavelength at the LRG redshift, i.e., $\lambda_{{\rm Mg}{\small \rm II}}\,(1+z_{\rm LRG}) = \lambda_{\rm QSO\,Line}\,(1+z_{\rm crit})$. We also mask the region blue-ward of C~{\small IV} emission lines

\begin{table}
\begin{center}
\caption{Number of pairs for each selection}
\label{tab_pairs}
    \begin{tabular}{lcc}
    \hline
    \multicolumn{3}{c}{QSO-LRG pairs with $r_p\leq1000$~kpc or $\leq2.5\,r_{vir}$: 1,712,316} \\ 
      & Removed & Remained \\  
    \hline
    \hline
    \multicolumn{3}{c}{Excluding Mg~{\small II}-associated absorbers} \\    
    &  $27,048$ ($\sim 2\%$) & $1,685,268$ \\
    \hline
    \multicolumn{3}{c}{Excluding region close to QSO C~{\small III} emission lines} \\    
    &  $134,935$ ($\sim 8\%$) & $1,550,333$ \\
    \hline
    \multicolumn{3}{c}{Excluding region blue-ward of C~{\small IV} emission lines} \\    
    &  $673,147$ ($\sim 40\%$) & $877,186$ \\
    \hline
    \end{tabular} 
\end{center}
\end{table}

Table~\ref{tab_pairs} summarizes the number of pairs for each selection criterion. 
The final sample consists of 877,186 LRG-QSO pairs from which we can detect Mg~{\small II} absorbers in the quasar spectra and study their links to the LRGs. 
We note that this sample is currently the largest dataset available for CGM studies, well exceeding $\sim 200$k LRG-QSO pairs used in previous analyses \citep[e.g.,][]{Anand2021}\footnote{This is the number of pairs obtained by cross-matching the SDSS DR16 QSO catalog with the Wisconsin PCA passive-galaxy sample that were used in \citet{Anand2021}, and by applying the same cuts listed in Table~\ref{tab_pairs}.}  

\subsubsection{Detecting the Mg~{\small II} absorption lines}
To detect absorption lines in the background quasar spectra, we use the automatic pipeline developed in \citet{Chang2024}. In the following, we summarize the main procedure of the pipeline.

We first use the Non-negative Matrix Factorization (NMF) method \citep{Lee1999,Zhu2016} to estimate the quasar continuum 
\citep{Zhu2013b}, along with a median filter of 85 pixels \citep{DESI_EDR_MgII} to remove small-scale fluctuations on the normalized spectra.
With the normalized spectra, we perform the following steps to identify MgII absorption line candidates:
\begin{enumerate}
    \item Shift the normalized spectra of background QSOs to the rest-frame of the foreground LRGs.

    \item Convolve the normalized spectra with a Mg~{\small II} doublet profile based on SDSS composite spectra \citep{Lan17} within a velocity window from $-1000$ to $1000~{\rm km/s}$ around the center of Mg~{\small II} wavelength (matched filter method).  
    
    \item Apply the same matched filter method to the spectral error arrays to obtain the uncertainty of the convolved spectra, retaining only the spectra with at least one pixel having detection significance greater than 2. 
    
    \item Identify the velocity of the pixel with the maximum detection significance from the remaining convolved spectra, considering that velocity as the central velocity of Mg~{\small II} absorber candidate. 
      
    \item Fit the original normalized spectra at the central velocities of Mg~{\small II} absorber candidates with a double Gaussian profile to characterize their absorption lines. 

    \item Select only Mg~{\small II} absorber candidates with SNR of $W_{0,\lambda2796}\geq 3$ and SNR of $W_{0,\lambda2803}\geq 2$ as detected Mg~{\small II} absorption lines. 
\end{enumerate}

We detected $13,712$ Mg~{\small II} absorption lines from the 877k LRG-QSO pairs using the above procedures, yielding a detection rate of $\sim 1.6\%$, consistent with the expectation that most pairs have large separations where $f_c$ is very low (see Section \ref{covering}).
From these detections, We obtain the properties of these absorbers, including their rest equivalent widths $(W_{0, \lambda 2796, 2803})$, line widths ($\sigma$), and central velocities, as well as their corresponding uncertainties. 
Figure~\ref{pair_fitting_example} shows an example of Mg~{\small II} absorption along a background-quasar sightline with Mg~{\small II} absorption line.

To avoid possible contamination from other absorption species, which accidentally fall within the search window and mimic the signals of Mg~{\small II} absorption lines, we remove absorption-line detections with the doublet line ratio ($R=W_{0,\lambda2796}/W_{0,\lambda2803}$) not being consistent with the expected values. 
More specifically, we calculate the line ratio distribution as a function of $W_{0,\lambda2796}$, taking the DESI EDR Mg~{\small II} absorber catalog \citep{DESI_EDR_MgII} as a reference. 
For a given $W_{0,\lambda2796}$, we exclude detections with $R$ values outside the empirical $3\sigma-$equivalent interval of the reference $R$ distribution. 
This leads to a removal of $263 (\sim 1.9\%)$ detected systems. 
To further identify potential contamination by Fe~{\small II}, for each detected absorption line, we assume that it is one of the Fe~{\small II} transitions at $\lambda2344$, $\lambda2374$, $\lambda2383$, $\lambda2586$, or $\lambda2600$. 
Based on this assumption, we check whether a corresponding Mg~{\small II} absorber exists at the expected longer wavelength. 
If an Mg~{\small II} absorber is detected at that position, we discard the originally detected systems. However, some of these contaminated lines may be a blend of genuine Mg~{\small II} and Fe~{\small II} lines.
Therefore, if we also identify a corresponding Fe~{\small II} line at a shorter wavelength than the originally detected Mg~{\small II} line, we retain the detection.
These steps additionally remove $391\, (\sim 2.9\%)$ features considered Fe~{\small II} contamination. 


\subsection{Stellar mass estimation} \label{stellar} 
To estimate the stellar masses of DESI LRGs for the pairs selected from Section~\ref{selecting_pair}, we perform the spectral energy distribution (SED) fitting using Code Investigating GALaxy Emission \citep[\texttt{CIGALE}][]{CIGALE}. 
We use the photometric fluxes provided by the DESI Legacy Imaging Surveys \citep{Dey2019} and the Wide-field Infrared Survey Explorer \citep[WISE, ][]{Wright2010} in five bands, $g$, $r$, $z$, W1, and W2, in the SED fitting. 
For stellar populations, we apply the \citet{BC2003} model and adopt the \citet{Chabrier2003} initial mass function. 
For all the other \texttt{CIGALE} modules applied in our fitting, the parameter values we use are shown in Appendix~\ref{cigale}. 
Most of the parameter settings are based on \citet{Siudek2024} with minor adjustments to better account for massive galaxies. 
The estimated stellar mass distributions are shown in Figure~\ref{stellar_dist}, with a median value of $\log_{10}\,{\rm M_*}/{\rm M_\odot} = 11.275$. 
We note that we also obtain other physical properties for the LRGs with \texttt{CIGALE}, such as star formation rate (SFR) and rest-frame color. 

\subsection{Halo mass and virial radius estimation} \label{virial}
Galaxies with different stellar mass reside in dark matter halos with different mass and sizes \citep[e.g.,][]{Conroy2009}. At fixed impact parameters in physical space, we may probe the gas corresponding to different scales in terms of the dark matter halos. 
To investigate the gas at different impact parameter relative to the halo size, we estimate the virial radius for each LRG based on its halo mass derived with Bayesian inference. 
Namely, we obtain the probability distribution of halo masses given an observed stellar mass by combining the likelihood of observing certain stellar masses for a given halo mass with the prior probability of that halo mass occurring in the universe \citep[e.g.,][]{Behroozi2010, Leauthaud2012, Moster2013}. 
The likelihood function is constructed using the stellar-to-halo mass relation (SHMR) from \citet{Moster2013}, assuming a Gaussian distribution for the observed stellar masses (data) given a possible halo mass (model). 
The mean and variance of the Gaussian correspond to the expected average stellar mass from the SHMR and the quadrature sum of the intrinsic stellar mass scatter and observational errors, respectively \citep{Leauthaud2012, Behroozi2013, Moster2013}.
The prior probability is constructed using the halo mass function (HMF) of \citet{Tinker2008a}. 

Given the observed stellar mass for each LRG, we determine its most probable halo mass, defined as the weighted average of the posterior distribution of the halo mass.
The virial radius can then be estimated using the analytic formula from \citet{Bryan1998}.
Figure~\ref{stellar_dist} illustrates the distributions of the estimated halo mass and virial radius (middle and right panels). 
The median values of the halo masses and virial radii for our samples are $\log_{10}\,{\rm M_{halo}}/{\rm M_\odot} = 13.1$ and $r_{\rm vir} = 416$~kpc, respectively. 
The median halo mass in our sample is lower than that inferred from halo occupation distribution (HOD) analyses, which yield a mean halo mass of $\log_{10}\,{\rm M_{halo}}/{\rm M_\odot}\sim13.2$ at $0.6<z<0.8$ \citep{Yuan2024}. 
In \citet{Yuan2024}, the mean halo mass is calculated as an N(M)-weighted average across mass bins, so more massive halos that host more satellite galaxies contribute greater weight.

\begin{figure} 
\begin{center}
\includegraphics[width=0.9\linewidth]{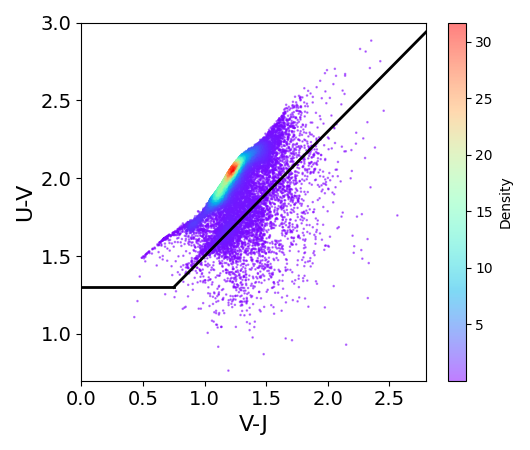}
\end{center}
\caption{U-V and V-J color-color distribution of DESI LRGs. The color bar represents the source density computed using a kernel density estimate (KDE). The thick black lines show the \citet{Whitaker12} color cut separating star-forming and quiescent galaxies.}
\label{UVJ_diagram}
\end{figure}

\subsection{Exclusion of blue and SFR-enhanced LRGs} \label{select_blue}
We find that a fraction of DESI LRGs exhibits relatively high SFR with bluer colors. 
To minimize potential bias of this blue population in our analysis, we exclude the blue galaxies using the rest-frame colors $U-V$ and $V-J$ (UVJ diagram) obtained from \texttt{CIGALE}. 
Figure~\ref {UVJ_diagram} shows the color-color distribution of DESI LRGs in the UVJ space. 
The red and blue LRGs are separated according to the criteria of \citet{Whitaker12}: $U-V > 0.8 \times (V-J) + 0.7~{\rm and}~U-V > 1.3$. 
The majority of DESI LRGs are quiescent and red, occupying the upper-left region defined by the criteria, while $\sim 9\%$ of the LRGs are classified as blue and SFR-enhanced. 
After removing the LRGs in the lower-left region, our sample consists of $809,679$ LRG-QSO pairs and $12,213$ detected Mg~{\small II} absorption lines. A comparison between the CGM properties of the two LRG populations is discussed in Appendix~\ref{blue_vs_red}. 

\subsection{The DESI LRG-QSO Mg~{\small II} absorber sample} \label{absorber_sample}
Our final LRG Mg~{\small II} absorber sample includes $12,213$ detected absorption lines from $809,679$ LRG-QSO pairs. 
Hereafter, we define ``weak" absorbers as having $0.4 \leq W_{0,\lambda2796} < 1 \rm\, \AA$ and ``strong" absorbers as those with $W_{0,\lambda2796} \geq 1 \rm\, \AA$. 
To determine the intrinsic properties of these absorbers, such as the incidence rate around galaxies, we must take into account the completeness of the whole sample. 
That is, the fraction of genuine absorbers missed due to low spectral SNR, rather than their true absence. 

To correct for incompleteness, we apply weighting factors based on simulations of detection rates using mock Mg~{\small II} absorption profiles with varying $W_{0,\lambda2796}$ and redshifts under different spectral SNR (see \citeauthor{Chang2024}~\citeyear{Chang2024} for details). 
The mock profiles are composite spectra built by stacking MgII absorption lines detected in the DESI year 1 quasar sample \citep{Napolitano2024} with the pipeline developed by \citet{DESI_EDR_MgII}. 
We further introduce spectral SNR thresholds 
with $W_{0,\lambda2796}$ below the noise level, where the SNR of quasar spectra is insufficient and the completeness correction becomes unreliable. 
For weak absorbers, we only consider the detection from the spectra with $SNR\geq 6$, and for strong absorbers, we only consider the detection from that with $SNR\geq 2.5$. 
The threshold values are determined from the simulations such that using weights to correct for undetected absorbers is effective, with the correction factor $\lesssim10$ (See Figure 12 in \citet{Chang2024}). 

\begin{table}
\caption{Number of detected Mg~{\small II} absorption lines and pairs used to detect absorbers. The SNR cuts for weak absorbers with $0.4 \leq W_{0,\lambda2796} < 1~{\rm \AA}$ and strong absorbers with $W_{0,\lambda2796} \geq 1~{\rm \AA}$ are 6 and 2.5, respectively. Here we also show the number of all detected absorbers with a uniform SNR cut of 6 for gas kinematics studies.}
\begin{center}
    \begin{tabular}{ccc}
    \hline
    \hline
     & Absorbers & Pairs \\
    \hline
    The whole sample & 12,213 & 809,679 \\
    \hline
    Weak absorbers with SNR~${\geq 6}$ & 3,399 & 146,943 \\\
    Strong absorber with SNR~${\geq 2.5}$ &  4,671  & 398,759 \\
    \hline
    All absorbers with SNR~$\geq 6$ & 5,276 & 146,943 \\  
    \hline
    \end{tabular}
\label{tab:sncut_absorbers}
\end{center}
\end{table}

We summarize the total number of both our samples and sub-samples with spectral SNR cuts in Table~\ref{tab:sncut_absorbers}. 
The median redshift for our selected LRGs is $0.768$. 
These (sub-)samples are applied to study the CGM properties of DESI LRGs in Section~\ref{results}.

\begin{figure*} 
\begin{center}
\includegraphics[width=1\linewidth]{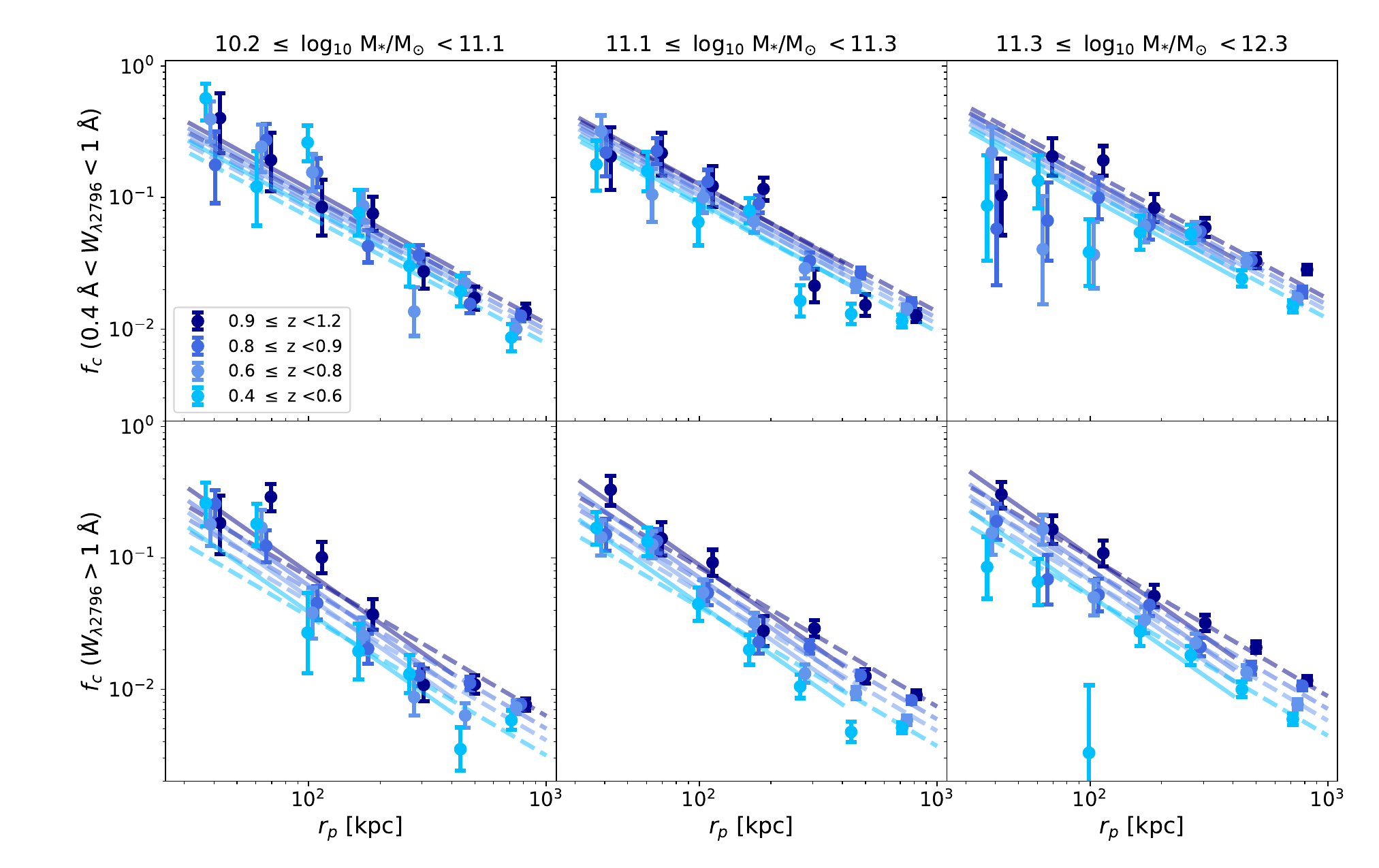}
\end{center}
\caption{Covering fraction as a function of impact parameter in physical space. Top: $f_c$ for absorbers with $0.4 \leq W_{0,\lambda2796}<1 \rm \, \AA$ (weak absorbers) Bottom: $f_c$ for absorbers with $W_{0,\lambda2796}\geq 1 \rm \, \AA$ (strong absorbers). The gas covering fractions for LRGs with stellar masses ranging from least to most massive are shown in the left to right panels. Colors transition from light blue to dark blue, indicating galaxy redshifts from lowest to highest. The errors are estimated based on binomial statistics. Solid lines and dashed lines illustrate the posterior median of the regression lines for $r_p \leq 400$~kpc and $r_p \leq 1000$~kpc, respectively.}
\label{cov_rp_physical}
\end{figure*}

\section{Results} \label{results}
\subsection{Gas distribution around DESI LRGs} \label{covering}
We begin with studying the radial distribution of the gas traced by Mg~{\small II} absorption lines around DESI LRGs. 
The covering fraction $f_c$, representing the probability of detecting absorbers around galaxies, provides a statistical measure of the gas distribution. 
In this work, we estimate the covering fraction with
\begin{equation}
    f_{c,obs} = \frac{\sum_{i}^{N_{abs}} w_{i}}{N_{\rm quasars}},
\end{equation}
where $w_{i}$ is the completeness weighting factor for each detected absorber $i$, $N_{abs}$ is the number of detected absorbers, and $N_{\rm quasar}$ is the total number of LRG-QSO pairs (or line of sight) used to search for absorbers. 
Appendix~\ref{pair_number} lists the number of pairs in each stellar mass, redshift, and $r_p$ bin.
We also account for the contribution from background absorbing structures not connected to DESI LRGs by measuring the average incidence rate in random sight-lines within the same velocity window (see Appendix~\ref{background} for more details). 
The background covering fraction $f_{c,b}$ is subtracted from the original observed $f_{c,obs}$. 
In the following, we denote the background-subtracted covering fraction as $f_c$. 
The $f_c$ is measured in both physical and dark matter halo space, as discussed in Section~\ref{fc_physical} and~\ref{fc_darkmatter}. 
The uncertainty of $f_{c}$ is estimated using the Wilson score interval at the $1 \sigma$ confidence level for binomial proportions.

\begin{table*} 
\begin{center}
\caption{
Posterior median as well as 68\% credible intervals of the regression parameters for the covering fraction $f_c$.}
    \begin{tabular}{cccccc}
    \hline
     \multicolumn{2}{c}{Physical Space} & $A$ & $\alpha$ & $\beta$ & $\gamma$ \\
    \hline   
    \hline
   $0.4 \leq W_{0,\lambda2796} < 1~{\rm \AA}$ & $r_p \leq 400$ kpc& $0.053\pm 0.014$ & $1.150\pm 0.382$ & $0.168\pm 0.096$ & $-0.994\pm 0.052$ \\
    & $r_p \leq 1000$ kpc & $0.041\pm 0.006$ & $1.254\pm 0.212$ & $0.450\pm 0.055$ & $-0.964\pm 0.026$ \\   
    \hline
   $W_{0,\lambda2796} \geq 1~{\rm \AA}$ & $r_p \leq 400$ kpc& $0.014\pm 0.003$ & $2.451\pm 0.330 $ & $0.304\pm 0.082$ & $-1.270\pm 0.042$ \\
    & $r_p \leq 1000$ kpc& $0.013\pm 0.002$ & $2.471 \pm 0.187$ & $0.360\pm 0.046$ & $-1.057\pm 0.022$ \\   
    \hline
     \multicolumn{2}{c}{Halo Space} & $A_{halo}$ & $\alpha_{halo}$ & $\beta_{halo}$ & $\gamma_{halo}$ \\
    \hline   
    \hline
   $0.4 \leq W_{0,\lambda2796} < 1~{\rm \AA}$ & $r_p \leq 0.5~r_{vir}$ & $0.041\pm 0.0169$ & $1.509\pm 0.544$ & $-0.763\pm 0.127$ & $-0.488\pm 0.096$ \\
    & $r_p \leq 2.5~r_{vir}$ & $0.007\pm 0.001$ & $2.654\pm 0.204$ & $-0.004\pm 0.050$ & $-0.939\pm 0.024$ \\   
    \hline
   $W_{0,\lambda2796} \geq 1~{\rm \AA}$ & $r_p \leq 0.5~r_{vir}$ & $0.003\pm 0.001$ & $3.571\pm 0.449 $ & $-0.484\pm 0.101$ & $-1.027\pm 0.065$ \\
    & $r_p \leq 2.5~r_{vir}$ & $0.002\pm 0.0002$ & $3.664 \pm 0.178$ & $-0.062\pm 0.043$ & $-1.037\pm 0.021$ \\   
    \hline
    \end{tabular}
\label{cov_fit_para}
\end{center}
\end{table*}

\subsubsection{Gas distribution in physical space} \label{fc_physical}

Figure~\ref{cov_rp_physical} shows the $f_c$ in two absorber strength bins for LRGs divided by stellar mass and redshift. 
The upper panels show the $f_{c}$ of weak absorbers with $0.4 \leq W_{0,\lambda2796} < 1,\text{\AA}$, while the lower panels show the $f_{c}$ of strong absorbers with $W_{0,\lambda2796} \geq 1,\text{\AA}$. 
The panels from left to right show $f_c$ of galaxies with different stellar masses. Redshifts of galaxies are indicated by the color gradients from light blue to dark blue. 
As shown in Figure~\ref{cov_rp_physical}, the $f_c$ of strong and weak absorbers decreases with $r_{p}$, consistent with the overall gas distribution traced by Mg~{\small II} absorbers around galaxies \citep[e.g.,][]{Zhu2014, Huang2016, Lan14, Lan2018, Lan2020, Huang2021, Anand2021}. 
Moreover, a positive correlation between $f_c$ and redshift 
is seen in Figure~\ref{cov_rp_physical}.

\begin{figure*} 
\begin{center}
\includegraphics[width=0.9\linewidth]{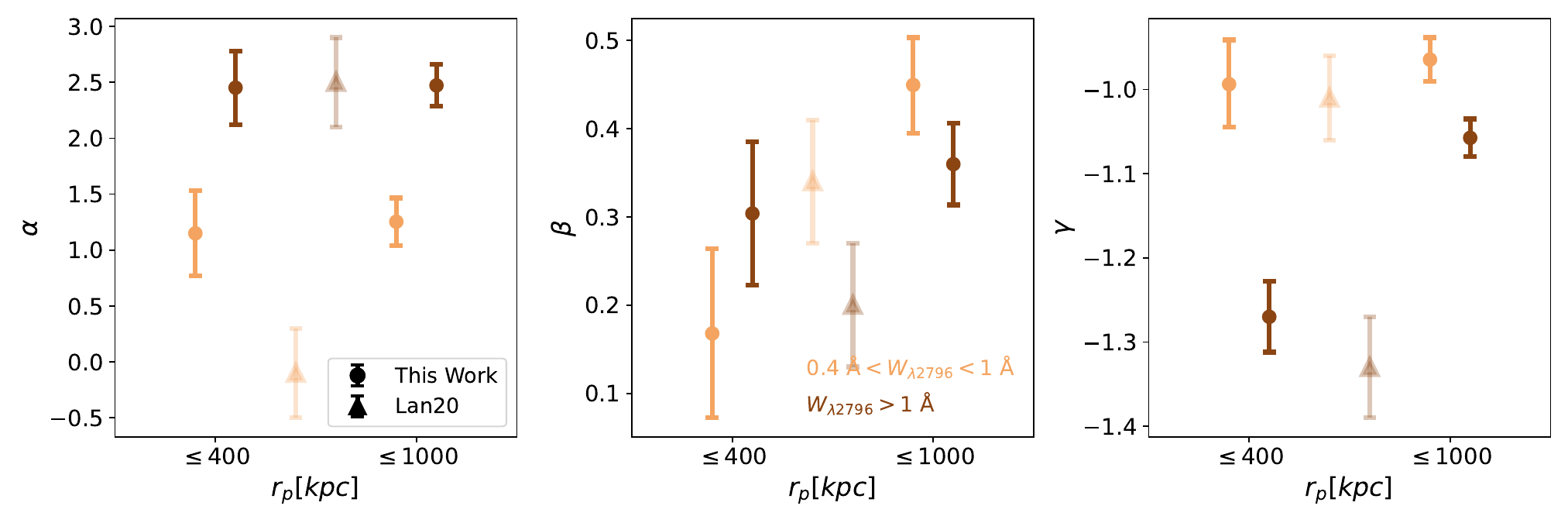}
\end{center}
\caption{Posterior median and 68\% credible interval of the regression parameters for the covering fraction $f_c$ in physical space. Left: parameter describing redshift dependence. Middle: parameter describing stellar mass dependence. Right: the slope of the gas distribution. Circles represent the MCMC-derived parameters obtained in this work, while triangles represent the best-fit parameters obtained in \citet{Lan2020} within $600$~kpc. Weak and strong absorbers are indicated by light brown and dark brown colors, respectively.}
\label{cov_para}
\end{figure*}

To quantify the correlation, we adopt an empirical power-law functional form from \citet{Lan2020} to characterize the $f_c$ measurements across different redshift and stellar mass bins,\begin{equation}
    \label{cov_eq}
    f_{c} = A\times(1+z)^{\alpha}\bigg(\frac{\rm M_*}{10^{11}\,\rm M_{\odot}}\bigg)^{\beta}\bigg(\frac{r_{p}}{100~{\rm kpc}}\bigg)^{\gamma},
\end{equation}
where A is the overall amplitude, $\alpha$, $\beta$, and $\gamma$ describe  the $f_{c}$ dependences on redshift, stellar mass, and $r_{p}$ respectively.
We perform global fittings of $f_{c}$ with two impact parameter ranges, (1) $r_p \leq 400$~kpc and (2) $r_p \leq 1000$~kpc with the best-fit profiles shown in Figure~\ref{cov_rp_physical} with solid and dashed lines respectively. 
We fit the $f_c$ with \texttt{emcee} \citep{emcee}, sampling the posterior distributions of the regression parameters with Markov chain Monte Carlo (MCMC). The regression parameter values of the $f_c$ are summarized in Table~\ref{cov_fit_para} and shown in Figure~\ref{cov_para}. 

The results show a number of trends:
\begin{itemize}
    \item  $f_c$ of both weak and strong absorbers evolve with redshift. The $\alpha$ parameter value of weak absorbers is $\sim1.3\pm0.21$ and that of strong absorbers is $\sim 2.5\pm0.19$ ($r_{p}<1000$ kpc), yielding $\sim 6\sigma$ and $13 \sigma$ detections for the redshift trend. 
    The best-fit $\alpha$ values of the two $r_p$ selections ($r_{p}<1000$ kpc and $r_{p}<400$ kpc) are consistent with each other, indicating such a trend is not primarily driven by the measurements at large scales (The left panel of Figure~\ref{cov_para}). 
    
    \item $f_c$ of strong absorbers ($\alpha\sim2.5$) evolve faster than $f_c$ of weak absorbers ($\alpha\sim1.2$).

    \item There is a stellar mass dependence with $\beta \sim 0.5 \pm 0.06$ and $\beta \sim 0.4 \pm 0.05$ within 1,000 kpc for weak and strong absorbers respectively, while such a trend is weaker within 400 kpc with $\beta \sim 0.2 \pm 0.10$ and $\beta \sim 0.3 \pm 0.08$ for weak and strong absorbers. (The middle panel of Figure~\ref{cov_para})
    

    \item For strong absorbers, $f_{c}$ decreases with increasing impact parameters with $\gamma$ being $-1.27\pm0.04$ within 400 kpc, which is steeper than the $\gamma$ value ($-1.06\pm 0.02$) when considering within 1000 kpc (The right panel of Figure~\ref{cov_para}). Such a difference is not detected for weak absorbers. 
\end{itemize}
The best-fit values obtained in \citet{Lan2020} are also shown in Figure~\ref{cov_para}. 
The redshift, stellar mass, and radial distribution trends are consistent with those reported in \citet{Lan2020} for passive galaxies. 
The only minor difference is that our results show a more pronounced redshift evolution for weak absorbers, which we will further discuss in Section~\ref{compare_previous}. 

\subsubsection{Gas distribution in halo space} \label{fc_darkmatter}

\begin{figure*} 
\begin{center}
\includegraphics[width=1\linewidth]{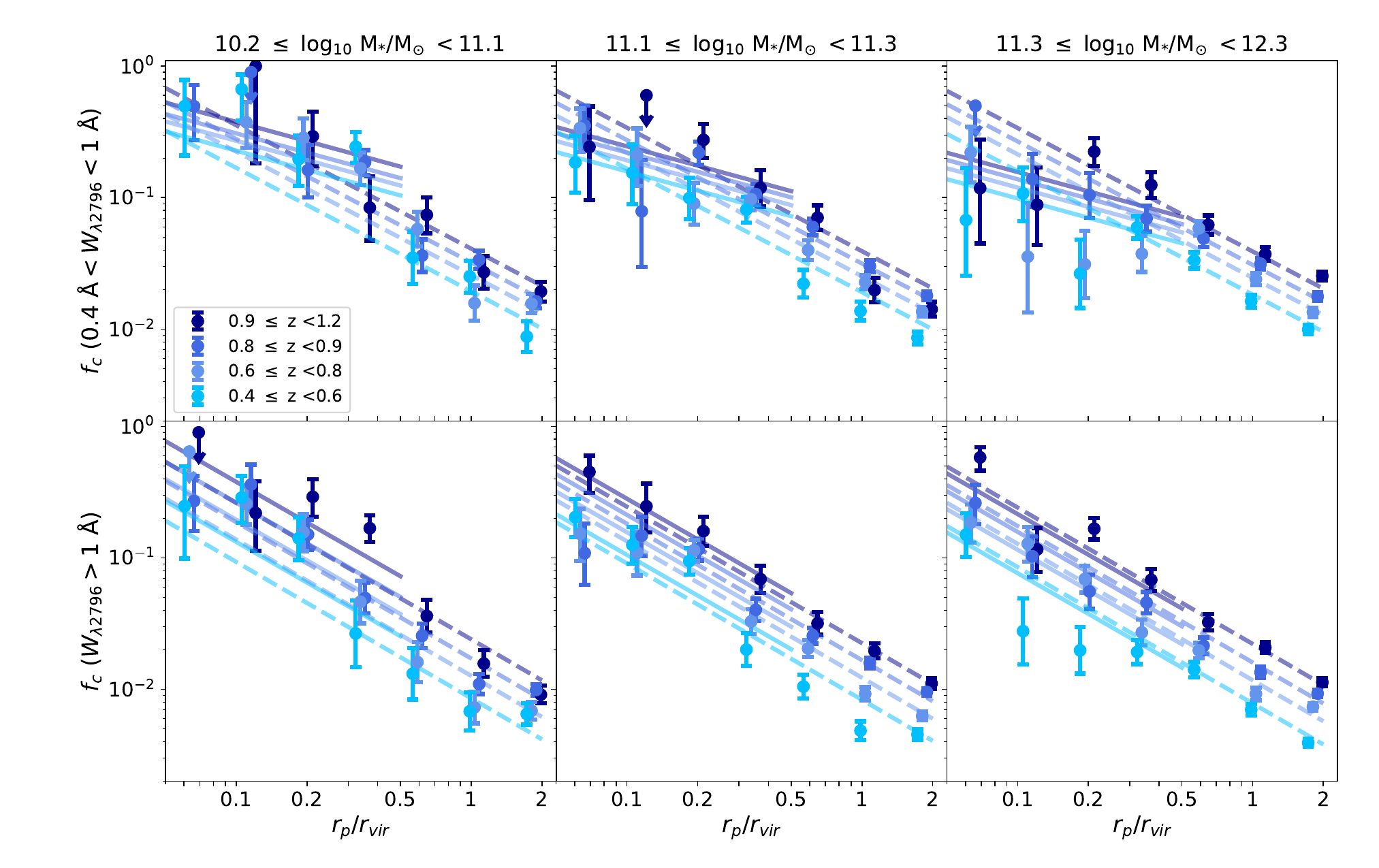}
\end{center}
\caption{Covering fraction as a function of impact parameter normalized by virial radius, for weak (upper panels) and strong absorbers (lower panels). From left to right, we show $f_c$ for LRGs with increasing stellar masses. The color gradient indicates redshifts of the galaxies. Errors are the confidence levels derived from binomial statistics. Non-detection bins (${\rm N_{det}}=0$) are shown with upper limits, determined with Wilson score statistics at a 3-sigma confidence level. Solid lines represent the posterior median of the regression lines for $r_p \leq 0.5~r_{vir}$, while dashed lines represent that for $r_p \leq 2.5~r_{vir}$.}
\label{cov_rp}              
\end{figure*}

Given that the size of halos depends on the stellar mass and redshift \citep[e.g.,][]{Conroy2009, Leauthaud2012, Behroozi2013, Moster2013}, a fixed physical impact parameter corresponds to a different distance relative to the virial radius
from the center of dark matter halos of galaxies at different redshifts. To better understand the gas distribution with respect to the size of the halos, in the following, we explore the gas distribution with the physical impact parameters normalized by virial radius of the dark matter halos.

The $f_{c}$ measurements are shown in Figure~\ref{cov_rp} with the x-axis indicating the physical impact parameter $r_{p}$ divided by the virial radius, $r_{vir}$, of the halos. 
The upper and lower panels show the measurements of weak and strong absorbers respectively with colors indicating the redshifts of galaxies. The left, middle and right panels show the results of three stellar mass bins listed in the figure.

Similarly to the measurements in physical space, we quantify $f_c$ in halo space following \citet{Lan2020} with a combined power profile,
\begin{equation}
    \label{cov_eq_halo}
    f_{c} = A_{halo}\times(1+z)^{\alpha_{halo}}\bigg(\frac{\rm M_*}{10^{11}\,\rm M_{\odot}}\bigg)^{\beta_{halo}}\bigg(\frac{r_{p}}{r_{vir}}\bigg)^{\gamma_{halo}},
\end{equation}
where $A_{halo}$ is the amplitude of the $f_{c}$, $\alpha_{halo}$, $\beta_{halo}$ and $\gamma_{halo}$ represent the dependence on redshift, stellar mass, and radius with respect to the size of halos. 
We perform the global fittings within $0.5~r_{vir}$ and $2.5\,r_{vir}$. 
The non-detection points (zero-detection bins, shown as upper limits in Figure~\ref{cov_rp}) are properly accounted for in the likelihood evaluation within \texttt{emcee}, with the upper limits derived using the Wilson score interval at a significance of $3\sigma$. 
Table~\ref{cov_fit_para} summarizes the regression parameters for $f_c$ in halo space. 

\begin{figure*} 
\begin{center}
\includegraphics[width=0.9\linewidth]{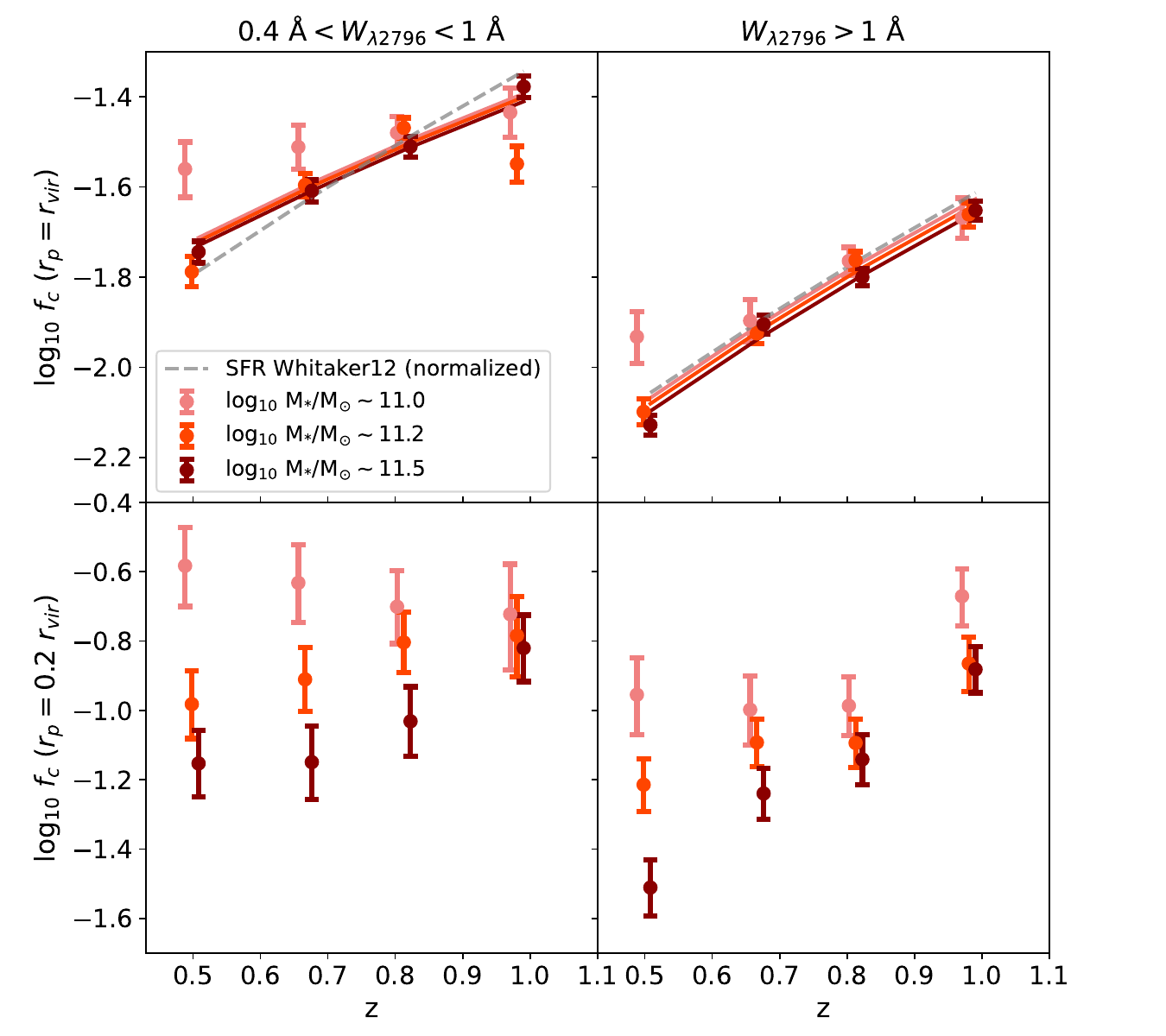}
\end{center}
\caption{MCMC-derived covering fraction as a function of redshift. Top: $f_c$ at $r_{vir}$. Bottom: $f_c$ at $0.2~r_{vir}$. The left and right panel shows the weak and strong absorbers, respectively. Plotted values correspond to the posterior median of the normalization parameter in each bin. The color scheme shows increasing stellar mass from light red to dark red. Solid lines represent the general trend of $f_c$ from the global fit, while the dashed line shows the SFR of galaxies as a function of redshift from \citet{Whitaker12}. The dashed lines are scaled for visual comparison and shown only in the top panels to illustrate halo-scale trends.}
\label{cov_para_norm}
\end{figure*}

Figure~\ref{cov_rp} shows several trends that are similar to those observed in the physical space:
\begin{itemize}
    \item Redshift dependence: 
    The results of the global fittings show that the redshift dependence $\alpha_{halo}$ values are $\sim2$ for weak absorbers and $\sim3$ for strong absorbers for both $r_{vir}<0.5$ and $r_{vir}<2.5$ scales. 
    To further illustrate the redshift evolution trend, we refit $f_c$ individually for each bin, covering $<2.5\,r_{vir}$ using $C\times(r_p/r_{vir})^{\gamma_{halo}}$ , with $\gamma_{\mathrm{halo}}$ fixed to the value reported in Table~\ref{cov_fit_para}. The posterior median of the normalization $C$ represents the characteristic $f_{c}$ at $1\times\,r_{vir}$ for each bin, as shown in Figure~\ref{cov_para_norm}. The trends from the global fits are shown by the solid lines, which are consistent with the general behavior of the values obtained from the individual fits. 
    
    From the individual measurements, we find that the correlations between $f_{c}$ and redshift for both weak and strong absorbers of LRGs with $10.2<\log_{10}\,{\rm M}_{*}/{\rm M_{\odot}}<11.1$ are not as strong as LRGs with higher stellar mass. By observing trends of $f_{c}$, we notice that such a stellar mass dependence is more prominent for the inner region of the halos $r_{p}<0.5\, r_{vir}$, motivating us to explore the redshift and stellar mass dependence in the inner regions of halos.  

    \item Mass dependence:
    To illustrate the trends of $f_{c}$ in the inner regions, we again fit the covering fraction individually with a fixed global slope from Table~\ref{cov_fit_para}, but restricted the fit to $r_p \leq 0.5 r_{vir}$. The MCMC-derived $f_c$ at $r_p = 0.2~r_{vir}$ is illustrated in the lower panel of Figure~\ref{cov_para_norm}. 
    We find that in addition to the redshift evolution, $f_{c}$ at $0.2\, r_{vir}$ depends on stellar mass. Galaxies with higher stellar mass have lower $f_{c}$ in the inner regions. This trend is also reflected in the stellar mass dependence global-fit $\beta_{halo}$ values with $r_{p}<0.5\, r_{vir}$ being $-0.76$ and $-0.48$ for weak and strong absorbers, respectively. This stellar mass trend is not observed at larger scales, indicating that there are possibly mechanisms associated with stellar mass suppressing primarily the $f_{c}$ of inner CGM. 

    We also observe that the redshift evolution of the inner CGM for different stellar masses is not entirely captured by the global fits. For example, 
    for weak absorbers (the left panel of Figure~\ref{cov_para_norm}), the $f_c$($r_p = 0.2~r_{vir}$) for the highest mass bin ($11.3<\log_{10}\,{\rm M}_{*}/{\rm M_{\odot}}$) decreases by a factor of 2 from redshift 1.1 to redshift 0.5, while the $f_c$($r_p = 0.2~r_{vir}$) for the low mass bin ($10.3<\log_{10}\,{\rm M}_{*}/{\rm M_{\odot}}<11.1$) does not decrease from redshift 1.1 to redshift 0.5. 
    The $f_c$($r_p = 0.2~r_{vir}$) of strong absorbers also shows this mass-dependent redshift evolution with an even stronger decreasing trend for the highest mass bin. 
    This indicates that the $f_c$($r_p = 0.2~r_{vir}$) of galaxies with higher stellar mass evolves faster than that of galaxies with lower stellar mass. 
\end{itemize}
These redshift and stellar mass evolution trends in $f_c$ offer novel information on the origins of the cool gas in massive halos. We will further discuss them in Section~\ref{discussion}. 
We note that $r_{vir}$, derived from stellar mass and redshift through halo mass estimates, may be affected by pseudo-evolution \citep[e.g.,][]{Liang2014}, which warrants caution when interpreting trends expressed in units of $r_p/r_{vir}$. We adopt $r_{vir}$ as a normalization for halo-space analyses because it is widely used as a physically motivated proxy for the overall halo extent, although it does not represent a uniquely defined physical boundary. Exploring alternative parameters that more accurately characterize the CGM in halo space is beyond the scope of this work. However, the redshift evolution of the covering fraction is detected in physical space, driving the overall redshift evolution trend in the halo space.  


\begin{figure*} 
\begin{center}
\includegraphics[width=1\linewidth]{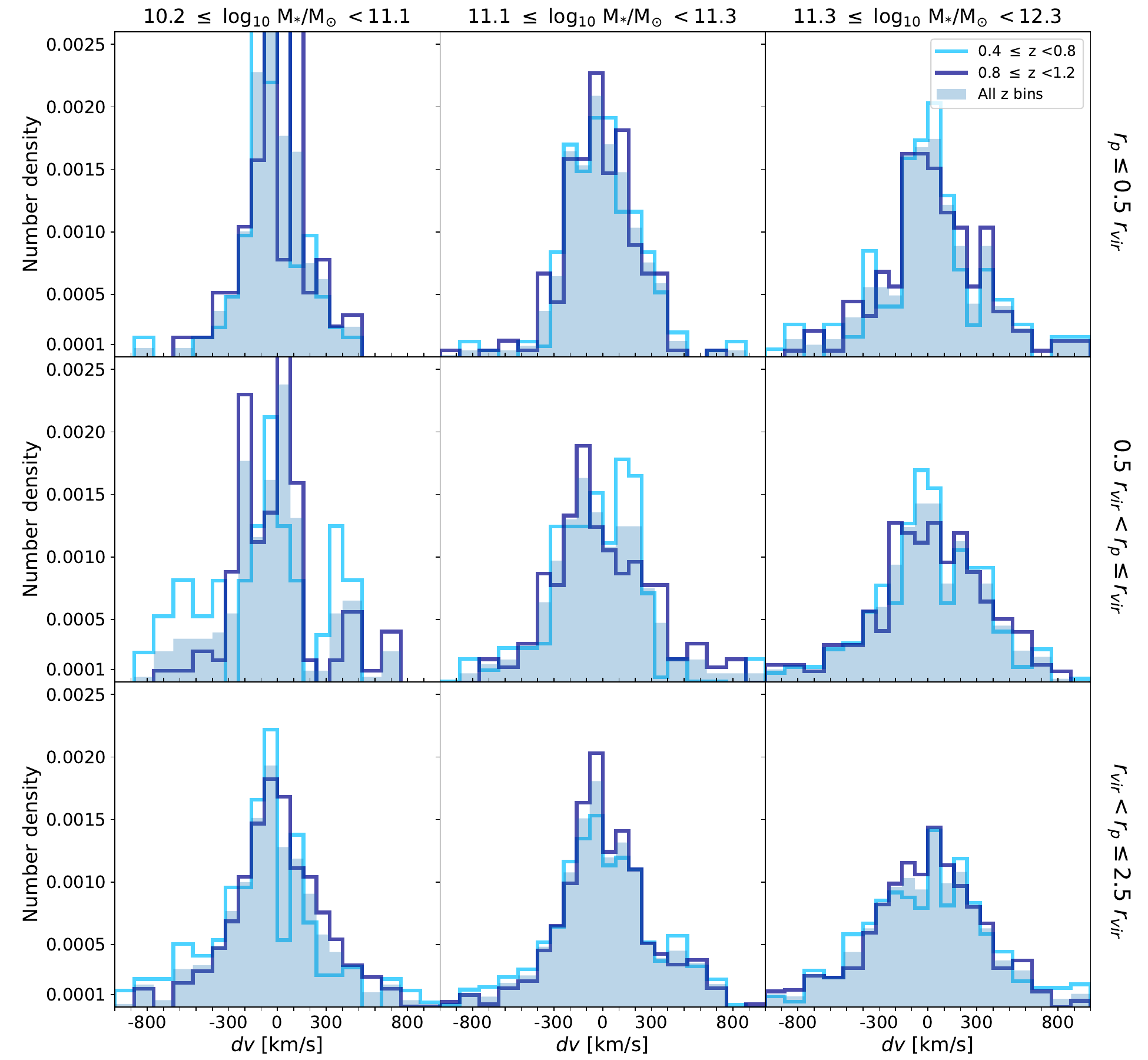}
\end{center}
\caption{Distributions of line-of-sight velocity difference ($dv$) between absorbers and LRGs within $r_p\leq0.5~r_{vir}$ (top panels), $0.5~r_{vir} < r_p \leq r_{vir}$ (middle panels), and $r_{vir} < r_p \leq 2.5~r_{vir}$ (bottom panels). Galaxies with the lowest to highest stellar masses are shown in the panels from left to right. The absorbers with different redshifts are indicated by light and dark blue curves, while the shaded region represents absorbers across all redshift bins. }
\label{vel_dist_hist}
\end{figure*}

\subsection{Gas kinematics} \label{losvelocity} 
We now investigate the kinematics of gas around DESI LRGs by measuring the line-of-sight (LoS) velocity difference, $dv$, between Mg~{\small II} absorbers and their associated galaxies. 
Figure~\ref{vel_dist_hist} shows the velocity distribution of absorbers associated with galaxies of different stellar masses and redshifts, within half of the virial radius (upper panels), between $0.5~r_{vir} < r_p \leq r_{vir}$ (middle panels), and between the virial radius and 2.5 $r_{vir}$ (lower panels). 
The observed distribution shown here is free from background contamination, as the estimated background contribution has already been subtracted.
Specifically, the expected number of background absorbers in each velocity bin was calculated by scaling the median random $f_{c,b}$ in velocity space (see Appendix~\ref{background}) with the total number of sightlines. 

\begin{figure*} 
\begin{center}
\includegraphics[width=1.\linewidth]{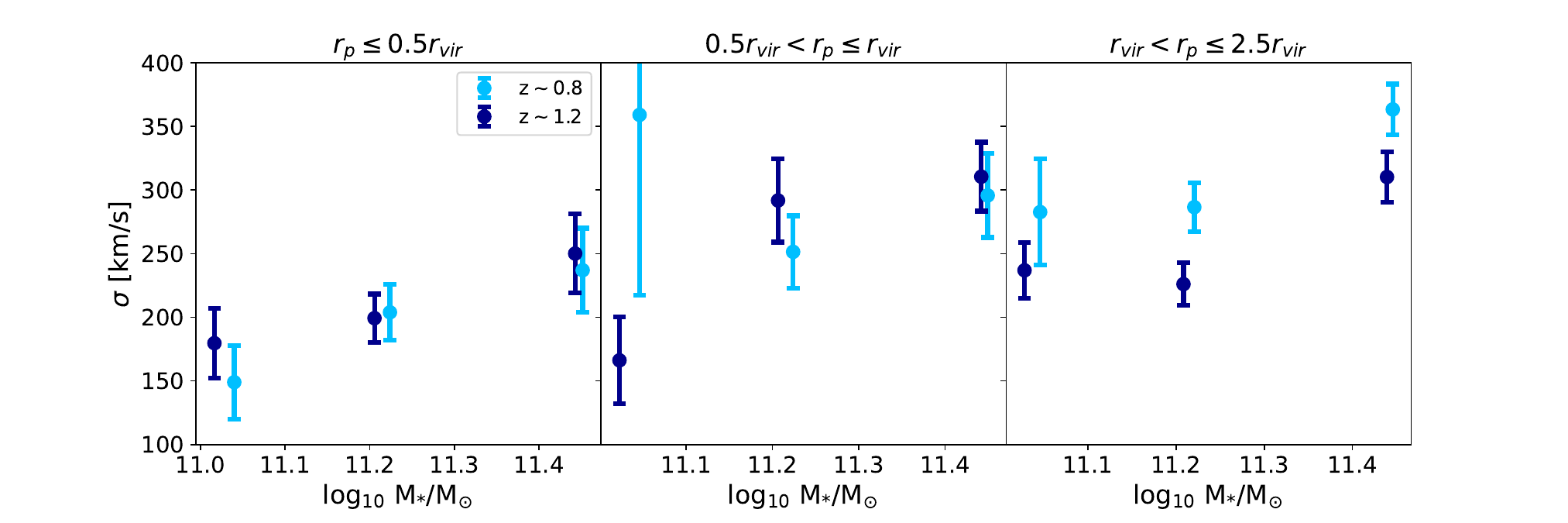}
\end{center}
\caption{Gas velocity dispersion as a function of redshift obtained from MAD with $r_p\leq 0.5~r_{vir}$ (left), $0.5~r_{vir}<r_p\leq r_{vir}$ (middle), and $r_{vir}<r_p\leq2.5~r_{vir}$. Light and dark blue represent redshifts of the galaxies. }
\label{mad_zdep}
\end{figure*}

As shown in Figure~\ref{vel_dist_hist}, a clear relationship is seen where more massive galaxies tend to have more high-velocity absorbers along the LoS, producing a broader distribution in velocity difference. 
This correlation becomes weaker for gas outside the halos.  
In Figure~\ref{mad_zdep}, we quantify the variation of gas velocity distributions with redshift, for each stellar mass and impact parameter bin, using the dispersion scaled from the median absolute deviation (MAD).  
The resulting MAD values indicate no significant differences between redshift bins. 
Furthermore, we apply the Kolmogorov–Smirnov (KS) test to evaluate the statistical consistency of the velocity distribution across redshifts, as detailed in the Appendix~\ref{veldisp_zdep}. 
No correlation between gas velocity and redshift is found. 
In the following, we measure the velocity difference by including all systems from $0.4<z<1.2$.

\subsubsection{Gas dispersion} \label{vel_disp}
We estimate the dispersion of LoS velocity difference for absorbers by measuring the velocity distribution with a combination of two Gaussian profiles \citep[e.g.,][]{Liang2014, Huang2016, Anand2021, Zu2021},

\begin{equation}
f_c (dv) =  A_n e^{\frac{-(dv)^2}{2\sigma_n^2}} ({\rm narrow})+ A_b e^{\frac{-(dv)^2}{2\sigma_b^2}} ({\rm broad}),
\label{fc_vel_gauss}
\end{equation}
where $A_n$ and $A_b$ are the amplitudes for the narrow and broad components and $\sigma_n$ and $\sigma_b$ are the corresponding dispersion values. 
A comparison of single- and double-Gaussian models is shown in Appendix~\ref{fit_velocity}, with the two-component model consistently favored ($\Delta{\rm BIC} \gtrsim 6$).
Since no significant redshift dependence is observed in the gas velocity, absorbers in each stellar mass and impact parameter bin were not further subdivided by redshift when estimating the velocity dispersion (see shaded blue regions in Figure~\ref{vel_dist_hist}). 

\begin{table*}
\caption{Posterior median as well as 68\% credible interval of the regression parameters for the velocity distribution. All absorbers with $0.4 < z < 1.2$ are included in the fitting, without redshift binning.
 }
\begin{center}
    \begin{tabular}{ccccc}
    \hline
    \hline
    &$\sigma_{n,0}$ & $\beta_n$ & $\sigma_{b,0}$ & $\beta_b$\\
    
    \hline
    All $r_p$ bins & $178\pm18$ & $0.038\pm0.140$ & $336\pm20$ & $0.254\pm0.050$\\
    $r_p \leq 0.5\, r_{vir}$ &$182\pm19$ & $-0.127\pm0.295$ & $258\pm83$& $0.644\pm0.331$\\
    $ 0.5 \, r_{rvir}< r_p \leq\, r_{vir}$&$214\pm63$ & $0.136\pm0.382$ & $377\pm110$& $0.058\pm0.561$\\
    $ r_{rvir}< r_p \leq\,2.5\,r_{vir}$& $176\pm27$ & $0.118\pm0.168$ & $369\pm30$& $0.188\pm0.065$ \\
    \hline
    \end{tabular}
\label{tab:dispersion_global}
\end{center}
\end{table*}

The gas dispersion can be inferred from the relative influence of the narrow and broad components. 
To better constrain both components, we perform a global fit across different stellar mass bins, parameterizing $\sigma$ as $\sigma_n=\sigma_{n,0}\times(\frac{{\rm M}_*}{10^{11}{\rm M}_\odot})^{\beta_n}$ and $\sigma_b=\sigma_{b,0}\times(\frac{{\rm M}_*}{10^{11}{\rm M}_\odot})^{\beta_b}$. 
This approach allows us to characterize the dependence of the dispersion on stellar mass in a global manner. 
Accordingly, Equation~\ref{fc_vel_gauss} involves a total of six fitting parameters $(A_n,\sigma_{n,0},\beta_n,A_b,\sigma_{b,0},\beta_b)$. 
We apply this global fitting procedure separately to each $r_p/r_{vir}$ bin, and also perform a combined fit over all distributions irrespective of impact parameter. 
Additional details of the global fitting are provided in Appendix~\ref{fit_velocity}. 
Table~\ref{tab:dispersion_global} shows the results of the global fits, suggesting that the gas dispersion ($\sigma_n$ and $\sigma_b$) does not depend on impact parameter. 
Therefore, we adopt the combined-fit results over all $r_p$ bins to estimate the velocity dispersion. 

\begin{table}
\caption{Fixed velocity dispersions of the narrow and broad components obtained from the global fit to all $r_p$ bins.}
\begin{center}
    \begin{tabular}{ccc}
    \hline
    \hline
    $\log_{10}\,{\rm M}_{*}/{\rm M}_{\odot}$ & $\sigma_n$ [km/s] & $\sigma_b$ [km/s] \\
    \hline
    $10.2  - 11.1$ $(\sim 11.0)$& $171\pm19$ & $339\pm20$\\
    $11.1  - 11.3$ $(\sim 11.2)$& $174\pm23$ & $378\pm24$\\
    $11.3  - 12.3$ $(\sim 11.4)$& $178\pm33$ & $429\pm33^{*}$\\
    \hline
    \end{tabular}
    \\[1ex] 
    \parbox{0.9\linewidth}{
    \small \textit{$^*$Note:} The median stellar mass of the most massive bin differs slightly between impact parameter ranges, resulting in a small difference in the value of $\sigma_b$. The median $\log_{10}\,{\rm M}{*}/{\rm M}{\odot}$ and $\sigma_b$ for the region within $r_{\rm vir}$ are 11.45 and 433 km/s, respectively, while those for $r_{\rm vir} < r_p \leq 2.5,r_{\rm vir}$ are 11.40 and 420 km/s.}
\label{tab:dispersion_massbin}
\end{center}
\end{table}

\begin{figure}
\begin{center}
\includegraphics[width=0.9\linewidth]{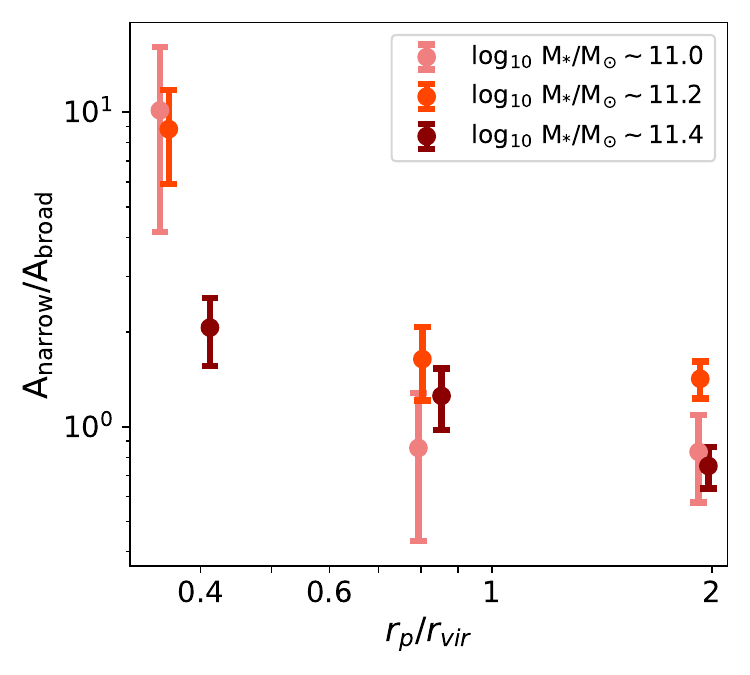}
\end{center}
\caption{Amplitude ratio between the narrow and broad components in the gas velocity distribution as a function of impact parameter. Colors from light red to dark red represent a progression toward higher stellar mass. }
\label{vel_disp_mass}
\end{figure}

The global fits yield the line widths of the velocity distributions ($\sigma_n$ and $\sigma_b$) as functions of stellar mass, as listed in Table~\ref{tab:dispersion_massbin}. 
We note that the contributions from the redshift uncertainty of the gas ($\sim 13~{\rm km/s}$) \citep{Chang2024} and the galaxies ($\sim 50~{\rm km/s}$) \citep{DESI_DR1} are subtracted to measure the intrinsic dispersion of the Mg~{\small II} absorbers. 
Based on the MCMC-derived parameters, the dispersion of the narrow component remains roughly constant with increasing stellar mass, while that of the broad component increases slightly. 
Fixing these values, we refit the distributions in each bin with Equation~\ref{fc_vel_gauss} to quantify the contribution of each component. 
Figure~\ref{vel_disp_mass} shows how the relative amplitudes of the narrow and broad components vary with impact parameter and stellar mass. 

Our analysis of velocity dispersion indicates several trends:
\begin{itemize}
    \item Radius dependence: As the impact parameter increases, the broad component becomes increasingly dominant. This trend is particularly pronounced in lower-mass galaxies, where the amplitude ratio between the narrow and broad components decreases by a factor of $\sim 10$, three times more than the reduction observed in massive galaxies (a factor of $\sim 3$). The enhanced prominence of the broad component at larger radii leads to an increase in the overall gas velocity dispersion, resulting in a radius-dependent trend in dispersion. 
    \item Mass dependence: At the inner region ($r_p \leq 0.5\, r_{\rm vir}$), the relative strength of the narrow to broad component decreases with stellar mass. The amplitude ratio declines from $\sim 10$ to $\sim 2$ as stellar mass increases from $\log_{10}\,{\rm M}*/{\rm M}\odot \sim 11.0$ to $11.4$. As a result, the overall gas dispersion is higher for absorbers associated with more massive galaxies. In contrast, no clear mass-dependent trend is observed at larger radii. 
\end{itemize}
Based on these observed radius and mass trends, we will examine the mechanisms responsible for the cool gas in LRGs, as discussed in Section~\ref{discussion}.

\subsubsection{Dependence on halo mass} \label{veldisp_halo}
\begin{figure}
\begin{center}
\includegraphics[width=0.9\linewidth]{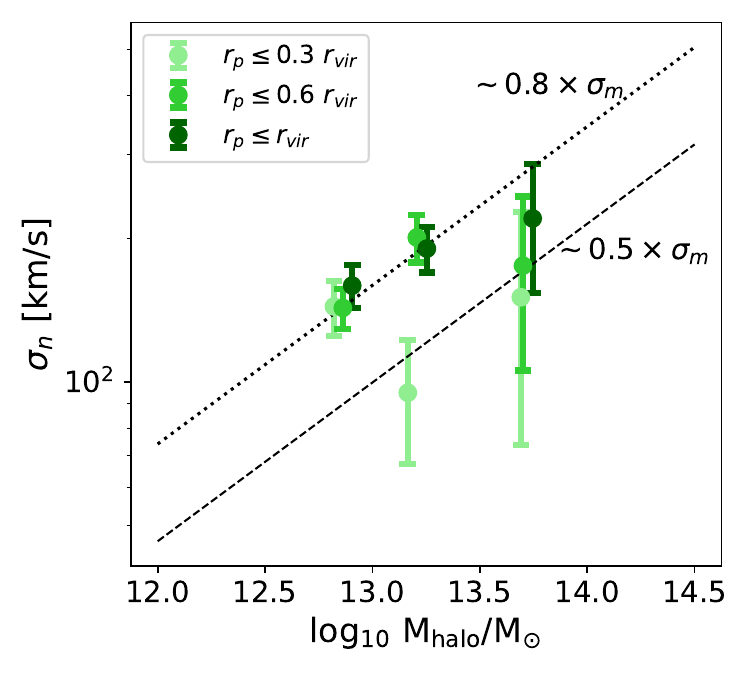}
\end{center}
\caption{Gas velocity dispersion for the narrow component as a function of halo mass. Color transition from light to dark green represents absorbers within 0.3 $r_{vir}$, 0.6 $r_{vir}$, and 1.0 $r_{vir}$, respectively. The dotted line shows the expected dark matter dispersion, while the dashed line shows 60\% of that value.}
\label{vel_disp_halo}
\end{figure}

To explore the connection between the kinematics of absorbers and their host dark matter halos, we examine the gas dispersion of absorbers within $r_{p}/r_{vir}<1$ as a function of halo mass. 
In this section, we focus on the narrow component, which dominates the velocity distribution at small radii. 
It is quantified using Equation~\ref{fc_vel_gauss}, with the line width of the broad component fixed to the values reported in Table~\ref{tab:dispersion_massbin}. 
The resulting correlation between the gas velocity dispersion of the narrow component and halo mass is shown in Figure~\ref{vel_disp_halo}. 
As shown in the figure, the gas dispersion of the massive halos ($M_{halo}>10^{13.5} M_{\odot}$) is approximately $50\%$ of the halo dispersion ($\sigma_m$) reported by \citet{ElahiVD}. 
which is consistent with previous studies \citep{Zhu2013b, Lan2018, Anand2021}. 
However, for the less massive bins in Figure~\ref{vel_disp_halo}, the gas dispersion values are $\sim 0.8~\sigma_m$, which are larger than the reported value for SDSS LRGs from \citet{Anand2021}. 

One possible reason for the higher dispersion observed, particularly among low-mass LRGs, is the difference in the methods used to estimate halo masses.
Our Bayesian approach yields systematically lower average halo masses than previous studies.  
At the high-mass end ($M_{\rm halo} > 10^{12.5}M_{\odot}$), halo mass cannot be reliably inferred from stellar mass using the SHMR alone, as the relation is not one-to-one \footnote{For a given halo mass, the SHMR yields a distribution of possible stellar masses rather than a single value.} due to the intrinsic scatter in stellar mass \citep[e.g.,][]{Tinker2017}.
Additionally, the number density of massive halos declines more steeply than that of lower-mass halos.
This, combined with the intrinsic scatter in the SHMR, can lead to overestimations of halo mass for massive galaxies, thereby shifting the data points in Figure~\ref{vel_disp_halo} toward the right.  
Applying Bayesian inference helps mitigate these issues and provides more robust halo mass estimates.

To assess the impact of different estimation methods, we recompute the halo masses using the empirical relation from \citet{Behroozi2010}, following \citet{Anand2021}.
This alternative approach results in significantly higher halo mass values by $\sim 0.5$ dex than those obtained from our Bayesian analysis.  
With these revised estimates, we recover a velocity dispersion–halo mass trend that aligns well with previous studies of LRGs. 
The data points originally associated with lower halo masses now shift to higher halo mass values, leading to a decrease in the inferred velocity dispersion at fixed halo mass.

We also estimate the velocity dispersion using stacked spectra, following the approach of \citet{Lan2018}.
Specifically, we stack all the spectra within the same halo mass and radial bins, and measure the line width of the resulting composite spectrum.
The velocity dispersion derived from the composite spectra is slightly lower than that estimated from individual spectra, particularly for low-mass galaxies, but the difference remains within the uncertainties. 

Overall, our measurements indicate a decline in the velocity dispersion, relative to the halo dispersion, as halo mass increases, with the most massive halos exhibiting values roughly half of the halo dispersion. 
This suppression in gas dispersion is largely consistent with previous studies \citep{Zhu2014, Lan2018, Anand2021} for passive galaxies. 
However, for less massive LRGs, we find dispersions comparable to the halo dispersion. 
This contrast highlights the need to examine a broader range of galaxy populations. 
Future studies focusing on high-mass star-forming galaxies or low-mass quiescent galaxies may offer a deeper understanding of this phenomenon.

\subsection{Mass of Circumgalactic Neutral Hydrogen}
In this section, we quantify the typical amount of cool H~{\small I} gas surrounding DESI LRGs using the covering fraction derived in Section~\ref{covering}. 
The mass of neutral hydrogen in the CGM ($M_{\rm H~{\small I}}$), as traced by Mg~{\small II} absorbers, can be estimated with 
\begin{equation}
    M_{\rm H~{\small I}}(<r_{vir}) \sim 2\pi\, m_{\rm H} \int_{0.02r_{vir}}^{r_{vir}} 	\hat{N}_{\rm H~{\small I}}\, f_{c}(r_{p})r_{p}dr_{p},
\end{equation}
where $\hat{N}_{\rm H~{\small I}}$ is the empirical relation between $W_{0,\lambda2796}$ and $N_{HI}$ from \citet{Lan17},
\begin{equation}
    \hat{N}_{\rm H~{\small I}} = 10^{18.96}\left(\frac{W_{0,\lambda2796}}{1 \,  \rm \AA}\right)^{1.69}(1+z)^{1.88} \, \rm cm^{-2}.
\end{equation}

\begin{figure}
\begin{center}
\includegraphics[width=1.\linewidth]{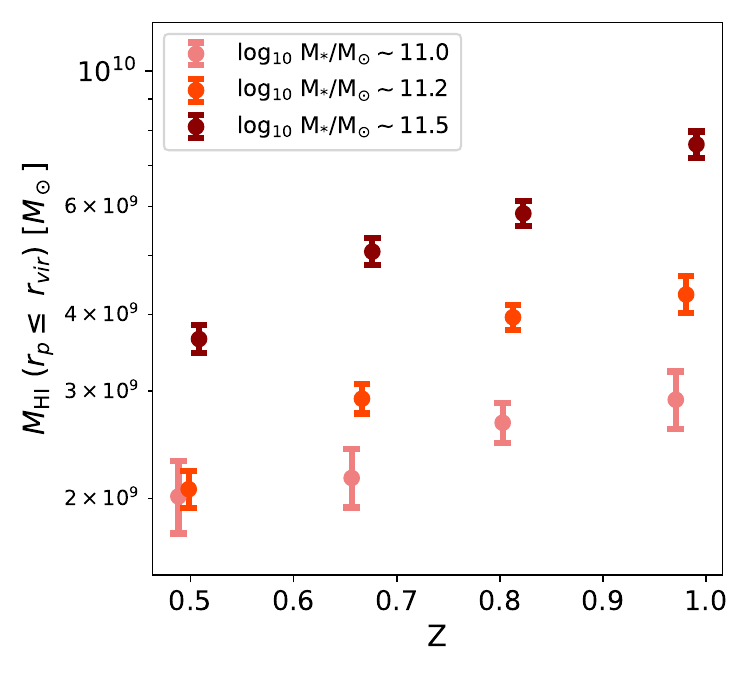}
\end{center}
\caption{Neutral Hydrogen mass within the halos as a function of redshift. Color gradients indicate the stellar masses of galaxies.}
\label{fig_himass}
\end{figure}

For weak absorbers, we take the median value of our MgII absorbers $W_{0,\lambda2796}\sim 0.65 \rm \,\AA$ and for strong absorbers, we take the median value $W_{0,\lambda2796}\sim 1.5 \,\rm \AA$. 
The lower integration limit of $0.02~r_{vir}$ is adopted to exclude gas associated with the interstellar medium of galaxies.
Figure~\ref{fig_himass} shows the abundance of H~{\small I} gas in the CGM of LRGs across different redshifts and stellar masses. 
The H~{\small I} mass within halos increases with both redshift and stellar mass, in agreement with \citet{Lan2020}. 
We note that systematic uncertainties are not included in the plot. 
Taking into account the redshift-dependent uncertainty in Equation 2 of \citet{Lan17}, systematics are estimated to be $\sim 0.1\,{\rm dex}$. 

We estimate that the enclosed $M_{\rm H~{\small I}}$ within the halos of LRGs is $\sim 3.5 \times10^{9}~{\rm M_\odot}$, given the median redshift and stellar mass of our sample. 
Clearly, a non-negligible amount of cool neutral gas was detected around massive galaxies, which has also been reported in the literature \citep{Zhu2013b, Huang2016, Lan2018, Lan2020, Anand2021}. 
The inferred $M_{\rm H~{\small I}}$ corresponds to fractions of $\sim 10^{-3.5}$ relative to the average halo masses of LRGs, in agreement with \citet{Lan2018}. 
We note that the H~{\small I} mass inferred here is larger than that derived from direct H~{\small I} absorption measurements \citep[e.g.,][]{Berg2019, Dutta2026}, due to the fact that strong Mg~{\small II} absorption preferentially traces high column density gas  with $N_{HI} \gtrsim 10^{19} cm^{-2}$ \citep[e.g.,][]{Rao17,Lan17}. Together with its covering fraction, this population traces the amount of H~{\small I} mass shown in Figure~\ref{fig_himass}.


Following \citet{Lan2018}, we quantify the cosmic mass density of neutral hydrogen contributed by the CGM of DESI LRGs. 
To do so, we assume that all massive galaxies with ${\rm M}_* \geq 10^{11}~{\rm M}_\odot$ share similar gas profiles. 
Applying the number density of $n \sim 10^{-2.9}~{\rm Mpc}^{-3}$ for quiescent galaxies from the stellar mass function in \citet{Moustakas13}, we estimate the cosmic mass density of H~{\small I} around LRGs as  
\begin{equation}
    \Omega^{\rm LRGs}_{\rm H~{\small I}} = \frac{n \times M_{\rm H~{\small I}}}{\rho_{\rm crit}} \sim 3.3 \times 10^{-5},
\end{equation}
where $\rho_{\rm crit}$ is the present-day critical density. 
As suggested by \citet{Lan17}, we adopt $\Omega^{\rm Mg~{\small II}}_{\rm H~{\small I}} \sim 1.5\times10^{-4}$ as the total mass density of neutral hydrogen traced by Mg~{\small II} absorbers at $z = 1$. 
These results indicate that the CGM of DESI LRGs contributes $\sim 22\%$ of the total neutral hydrogen mass density traced by Mg~{\small II} absorbers. 

\subsection{Summary of the observed results} 
In summary, our results show several trends in the properties of DESI LRG Mg~{\small II} absorbers:
\begin{itemize}
    \item {\bf Redshift trend: } The covering fraction of MgII absorption around LRGs is positively correlated with redshift.
    \item {\bf Radius trend: } The velocity dispersion is positively correlated with the impact parameter, reflecting the shift from narrow- to broad-component dominance at larger radii. 
    \item {\bf Mass trends at small radii: } The covering fraction anti-correlates with stellar mass, while the velocity dispersion positively correlates with stellar mass. Within the central halo region, the relative contribution of the narrow component decreases with stellar mass.  
\end{itemize}
These trends highlight key clues to the origins of the cool gas in massive halos. The implications of these findings are explored in Section~\ref{discussion}. 

\section{Discussion} \label{discussion}

\subsection{Redshift evolution of Mg~{\small II} absorbers around LRGs} 
In Section~\ref{fc_darkmatter}, we show that $f_c$ of MgII absorbers around LRGs increases with redshift. 
This trend is similar to the global redshift evolution of MgII absorbers, especially the strong absorbers as found in previous studies \citep[e.g.,][]{Prochter06, Menard11, Matejek12, Zhu2013b, Chen2017, Lan2020}. 
These results indicate that the cosmic number density of strong MgII absorbers follows the redshift evolution of the cosmic star formation rate density, suggesting a global connection between star formation and strong MgII absorbers. 
Specifically, the link may be driven by SFR-induced, metal-enriched outflows that populate the halo with cool, low-ionization gas, while other processes, such as gas accretion, cooling, and AGN feedback, can also regulate the cool gas content, reflecting a common dependence on the overall baryon cycle. 

To test whether this global link is also reflected in the local distribution of Mg~{\small II} absorbers around LRGs, in Figure~\ref{cov_para_norm}, we show the redshift evolution of SFR (grey line) from \citet{Whitaker12} at $\log_{10}\,{\rm M_*}/{\rm M_\odot} \sim 10$ and compare it with the $f_c$ of strong absorbers. 
Notably, the $f_c$ evolution resembles that of the SFR, suggesting that the evolution of the cool gas covering fraction around massive passive galaxies may follow the global trend of star formation activity even though the central galaxies have little or no star-formation activity. 
This suggests that the presence of strong Mg~{\small II} absorbers may not directly trace the ongoing star formation activities of the central galaxies. Instead, the presence of strong Mg~{\small II} absorbers might connect to the total star-formation rate in the halo, i.e. the star-formation of satellite galaxies.

We further explore the evolution of satellite populations within massive halos and compare it with the redshift evolution in the Mg~{\small II} covering fraction.
Specifically, the total fractions of star-forming and quiescent satellites in massive clusters are derived by integrating the stellar mass functions of cluster galaxies in the literature. 
At high redshift ($z \sim 1$), \citet{vanderburg2013,vanderburg2020} reported a star-forming satellite fraction of $\sim 0.3$, while at a lower redshift ($z \sim 0.6$), the fraction is $\sim 0.1$ \citep{vanderburg2018}. 
This threefold reduction in the star-forming satellite fraction aligns well with our measured decline of $\sim 0.45$ dex in the Mg~{\small II} covering fraction at $r_{vir}$, as seen in Figure~\ref{cov_para_norm}. 
This agreement is consistent with the interpretation that 
the incidence rate of Mg~{\small II} absorbers around LRGs may correlate with the abundance of 
star-forming satellites. 

The connection between strong Mg~{\small II} absorbers and halo-wide star formation is also supported by the findings of \citet{Lan2020}, who showed that 
the contribution of the total gas cross section around star-forming satellite galaxies can explain $>50\%$ of the MgII $f_{c}$ around massive passive galaxies. 
Recent simulation results by \citet{Staffehl25} also demonstrate that the cool gas associated with star-forming satellite galaxies contributes to a large fraction of observed cool gas in massive dark matter halos. 



\subsection{Mg~{\small II} absorbers in the inner region of LRG halos}
\subsubsection{Behaviors of $f_c$ at small radii}
We now discuss possible mechanisms that give rise to the stellar mass dependence of $f_c$ in the inner CGM as shown in the lower panels of Figure~\ref{cov_para_norm}. 
As previously discussed, the redshift evolution of $f_c$ may be linked to the redshift evolution of star-forming satellite galaxies in these systems. The correlation between $f_c$ and stellar mass in the inner regions can also be attributed to the association of the radial distribution of star-forming satellite galaxies. As more massive central galaxies tend to reside in higher density halos and environments \citep[e.g.,][]{White1978, Bryan1998, Kauffmann2004, Li2006, Yang2007, Peng2010}, which enhance ram-pressure stripping and therefore the removal of cool gas from satellites \citep{Gunn1972}, the fraction of star-forming satellite galaxies toward the inner regions is expected to decrease. Such a trend has been observed in galaxy groups and clusters in the local Universe \citep[e.g.,][]{Balogh1997, Weinmann2006, Wetzel2012, Woo13, Barsanti18}. 
We note that the stellar mass–dependent trends of $f_c$ at small $r_p$ may be a consequence of such mass dependence, leading to a drop in $f_c$ for more massive galaxies. 
New measurements at higher redshifts and as a function of redshift will be essential to further confirm the scenario. 

From a simulation perspective, more recent infalling satellites (i.e., those accreted most recently) are indeed more likely to retain their cool gas, as they have not yet experienced the cumulative effects of ram‑pressure stripping \citep[e.g.,][]{Rohr24, Staffehl25}. 
Consequently, such satellites might be predominantly located in the outer halo, where the ambient density is lower and stripping is less efficient. Observations of cool gas in clusters also reveal enhanced H~{\small I} depletion in denser environments toward their center \citep[e.g.,][]{Yoon2013, Yoon2017, Burchett18}, further supporting this scenario. 

Another possible mechanism for the lower $f_{c}$ at small radii around massive galaxies is associated with the thermal properties of the hot gas. 
Precipitation driven by thermal instability has been proposed as a plausible mechanism of producing cool gas in massive halos \citep[e.g.,][]{Werner2014, Voit2015a, Voit2017, Tremblay2018}.
In more massive halos, the elevated virial temperatures lengthen the gas cooling time, preventing the onset of thermal instability and consequently suppressing in-situ cooling and condensation \citep[e.g.,][]{Sutherland1993, Birnboim2003, Sharma2012, Voit2015b}. 
Moreover, feedback from AGNs may further suppress the accumulation of cool gas in the inner regions of massive galaxies (e.g., \citealt{Fabian2012}) and regulate the precipitation \citep[e.g.,][]{Voit2015b, Gaspari2020}. 
AGN-driven jets can inject mechanical energy into the surrounding medium, generating shocks and turbulence that heat the gas and suppress cooling and star formation \citep[e.g.,][]{McNamara2012, Heckman2014, Hardcastle2020, ARAA_simulations}. 
Overall, the thermal state of the hot gas is closely linked to halo mass and helps explain the reduced cool-gas content in the inner regions of massive halos. In this context, we note a noticeable flattening in the radial dependence of $f_c$ around $r_p \sim 200$ kpc (i.e., $\sim 0.4$–$0.7\,R_{\rm vir}$; Figures~\ref{cov_rp_physical} and~\ref{cov_rp}), which may be comparable to the expected cooling or precipitation radius of LRG-sized halos \citep{Voit2015a}, though this interpretation remains tentative.

The above scenarios can both explain the decline of $f_{c}$ in the inner CGM with increasing stellar mass. 
We note that a similar trend was also identified by \cite{Lan2020} and  \cite{Anand2021}. 
\citet{Lan2020} attributed the trend primarily to the most massive bin and reported a mass-dependent slope of $\sim 0.6$ for strong absorbers, which is consistent with our results in Section~\ref{fc_darkmatter}, while they found a shallower slope for weak absorbers. 
On the other hand, \citet{Anand2021} also suggested that the higher virial temperatures in more massive galaxies suppress the cooling of gas. 

\subsubsection{Behaviors of gas dispersion at small radii}
As discussed in Section~\ref{vel_disp}, the gas velocity distribution around LRGs can be described with two Gaussian components, suggesting the presence of at least two kinematic populations. 
For the narrow component, we argue that both satellite ram pressure stripping and precipitation scenarios could account for the observed Mg~{\small II} absorber velocity dispersion around massive galaxies \citep[e.g.,][]{Zhu2014, Huang2016, Lan2018}.
The velocity dispersion of gas stripped from satellites is decelerated by ram-pressure drag force, and high-velocity clouds are expected to be destroyed by hydrodynamic instabilities, resulting in velocity dispersion values lower than the velocity dispersion supported by the host dark matter halos \citep[e.g.,][]{Murray2004, Heitsch2009, Emerick2016, Armillotta2017}. 
Similarly, in-situ condensed clumps arising from local thermal instabilities and precipitation are expected to exhibit low initial velocities and small velocity dispersion of $\sigma\sim100~{\rm km/s}$ \citep[e.g.,][]{Gaspari2018, Fielding2020}. 
While gravity drives the infall of condensed clumps, hydrodynamic drag suppresses their bulk motion and damps velocity differences \citep[e.g.,][]{Voit2015b}, resulting in a small velocity dispersion among the surviving clumps consistent with observations of narrow components.  

On the other hand, the broad component may result from a combination of several mechanisms, including infalling gas associated with satellites, gas associated with galaxies in the nearby large-scale structure, and AGN feedback. 
These processes are expected to increase the velocity dispersion of absorbers around massive galaxies. 
As halos grow more massive, they tend to accrete more satellites and gas \citep[e.g.,][]{Gao2004, Wetzel2013} and exhibit stronger AGN feedback \citep[e.g.,][]{Best2005, Croton2006, McNamara2007}. 
Additionally, the lines of sight can intercept absorbers at large physical distances (but with small impact parameters) which follow the Hubble flow \citep[e.g.,][]{Zhu2014} and contribute to the observed broad component in the velocity distribution \citep[e.g.,][]{Turner2014, Huang2016}.  
This contribution of absorbers is expected to be more prominent in denser environments \citep[e.g.,][]{Turner2014}, where massive galaxies reside. 

In Section~\ref{vel_disp}, we also show that the amplitude ratio between the narrow and broad components anti-correlates with stellar mass.
Specifically, the narrow component becomes less prominent with increasing stellar mass, resulting in an anti-correlation between the gas dispersion and stellar mass at small radii. 
We suggest that this trend may be driven by less efficient precipitation and more dominant AGN feedback in more massive galaxies. 
Increased satellite quenching in massive halos, leaving less stripped gas in the central region, may also help explain the observed behavior. 
In addition, the increasing contribution of the broad component with stellar mass may be explained by enhanced accretion rate in more massive halos. 

Both satellite and precipitation origins could account for the observed mass dependence in the inner region.
We note that \citet{Anand2021} also observed a similar mass trend and attributed the correlation to greater gas accretion in higher-mass halos, originating from both the intergalactic medium and infalling satellites.

\subsection{Mg~{\small II} absorbers at Large Radii} 
There is a correlation between the gas dispersion and impact parameter as shown in Section~\ref{vel_disp}, indicating that the contributions of narrow and broad components depend on the impact parameters. Here we discuss possible scenarios that can explain such a trend. 
First, the abundance of the cool gas associated with the halos with low velocity dispersion (narrow component) decreases with distances from the center of halos, leading a decrease of the contribution of the narrow components. This can be due to the low precipitation efficiency at the outer region \citep[e.g.,][]{Voit2015b, Voit2017, Voit2021}. 
Second, the outskirts of halos are likely to host more accreting satellites and gas \citep[e.g.,][]{Tal2013, Wetzel2013, Hafen2019}, contributing to the broad component. 
Third, the contribution of the absorbers originated from the nearby large-scale structures will be more significant in the outer regions \citep[e.g.,][]{Tinker2008b, Prochaska2011, Zhu2014, Turner2014}. 
This interpretation is consistent with recent results from stacked Ly$\alpha$ absorbers by \citet{Dutta2024b}, who showed that broad line-of-sight velocity wings in the outer halo regions can be well described by the galaxy–absorber two-point correlation function.
Any of the proposed scenarios could lead to either a more prominent broad component or a diminished narrow component at a larger radius. 

\section{Conclusions} \label{conclusion}
We probe the cool gas around massive galaxies using $\sim 0.8$ million LRG-QSO pairs from DESI, with the gas traced by Mg~{\small II} absorption lines. Our sample size is about four times larger than the largest sample used in previous studies. 
We further refine the measurements by estimating halo masses and virial radii through a Bayesian framework that incorporates intrinsic SHMR scatter and the rapidly declining halo mass function, and by correcting for unrelated background absorbers using randomized galaxy–QSO pairs.

Leveraging this unprecedentedly large spectroscopic dataset and a more comprehensive analysis approach, we characterize the cool CGM of LRGs across a range of redshifts and stellar masses, measured at different impact parameters. 
With the DESI dataset, we are able to probe the CGM of massive galaxies at $z>0.8$, a regime that is not well covered by the SDSS LRG sample.
We also estimate the covering fraction in the inner halo (down to $0.2\,r_{\rm vir}$), and we measure the relative amplitudes of the narrow and broad velocity components as a function of stellar mass and impact parameter.
These gas properties provide indispensable information about how these massive galaxies evolve.
Our main results show that: 
\begin{itemize}
    \item The covering fraction is higher at higher redshift, with a redshift dependence $(1+z)^{\alpha}$ with $\alpha \approx 3.6$ for strong absorbers and $\approx 2.4$ for weak absorbers (Figure~\ref{cov_para_norm}). 
    \item The covering fraction is lower for more massive galaxies in the central region, with a mass trend $M_{*}^{\beta}$ with $\beta$ values ranging from $ \approx -0.8$ to $-0.6$ (Figure~\ref{cov_para_norm}). 
    \item The gas dispersion shows no dependence on redshift. It is higher for more massive galaxies in the inner region, with the relative strength between the narrow and broad components decrease from $\sim 10$ at $\log_{10}\,{\rm M_{*}}/{\rm M_\odot} \sim 11$ to $\sim 2$ at $\log_{10}\,{\rm M_{*}}/{\rm M_\odot} \sim 11.4$. Besides, the dispersion is higher at the outer halo, with the narrow-to-broad amplitude ratio decreasing by a factor of $\sim 3 - 10$ from $r_p\sim 0.3\,r_{vir}$ to $\sim 2\,r_{vir}$ (Figure~\ref{vel_disp_mass}). 
    \item Within $r_{vir}$, most massive LRG halos exhibit a velocity dispersion of $\sim 0.5\,\sigma_m$, while that of less massive halos is $\sim 0.8\,\sigma_m$ (Figure~\ref{vel_disp_halo}). 
    \item The total H~{\small I} mass, as traced by Mg~{\small II} absorbers around massive galaxies, is $\sim 3.5\times 10^9\,{\rm M}_\odot$, indicating that $\sim22\%$ of the Mg~{\small II} absorbers may be attributed to the CGM of massive galaxies (Figure~\ref{fig_himass}). 
   
\end{itemize}

These observations are consistent with a scenario in which most of the cool gas in massive galaxies originates from (i) infalling or stripped satellites and (ii) condensed clumps driven by in-situ thermal instability \citep[e.g.,][]{Huang2016, Chen2018, Lan2020, Staffehl25}. 
The similarity in the evolutionary trends of massive galaxy covering fraction and the galaxy star formation rate suggests a connection between cool gas and global star formation activity in halos. 
The reduced detection of cool gas along sightlines in the innermost regions of more massive galaxies further suggests quenching of cool gas in both central galaxies and their satellites, likely driven by more stable hot halo environments and enhanced gas stripping near the center. 
In the future, more extensive spectroscopic datasets, such as the full five-year DESI survey, will enable more detailed studies of the cool gas around massive galaxies. 
Such data can provide stringent tests for galaxy evolution models.
For example, the LoS velocity distributions place direct constraints on the kinematics of cool CGM gas in the simulation, offering a bridge between empirical measurements and theory \citep[e.g.,][]{Afruni2019}.
These constraints allow theorists to test feedback and accretion prescriptions in next-generation galaxy-evolution simulations \citep[e.g.,][]{Pakmor2023, Augustin2025} and to gain new insights into the origins of the cool gas and the processes that quench massive galaxies. 

\section*{Data Availability}
All data points shown in the figures are available in a machine-readable form on Zenodo \url{https://doi.org/10.5281/zenodo.17787068}.

\section*{Ackowledgments}
We thank Ming-Feng Ho and Paul Martini for their constructive comments that greatly helped improve the paper. YLC and TWL were supported by the National Science and Technology Council (MOST 111-2112-M-002-015-MY3, NSTC 113-2112-M-002-028-MY3), Yushan Fellow Program by the Ministry of Education (MOE) (NTU-110VV007, NTU-111V1007-2, NTU-112V1007-3, NTU-112V1007-4, NTU-112V1007-5) and MOE-115-YSFMS-0003-001-P2 (NTU 115V2020-1). YLC is supported by NSTC 113-2811-M-002-067-MY3.

This material is based upon work supported by the U.S. Department of Energy (DOE), Office of Science, Office of High-Energy Physics, under Contract No. DE–AC02–05CH11231, and by the National Energy Research Scientific Computing Center, a DOE Office of Science User Facility under the same contract. Additional support for DESI was provided by the U.S. National Science Foundation (NSF), Division of Astronomical Sciences under Contract No. AST-0950945 to the NSF’s National Optical-Infrared Astronomy Research Laboratory; the Science and Technology Facilities Council of the United Kingdom; the Gordon and Betty Moore Foundation; the Heising-Simons Foundation; the French Alternative Energies and Atomic Energy Commission (CEA); the National Council of Humanities, Science and Technology of Mexico (CONAHCYT); the Ministry of Science, Innovation and Universities of Spain (MICIU/AEI/10.13039/501100011033), and by the DESI Member Institutions: \url{https://www.desi.lbl.gov/collaborating-institutions}. Any opinions, findings, and conclusions or recommendations expressed in this material are those of the author(s) and do not necessarily reflect the views of the U. S. National Science Foundation, the U. S. Department of Energy, or any of the listed funding agencies.

The authors are honored to be permitted to conduct scientific research on I'oligam Du'ag (Kitt Peak), a mountain with particular significance to the Tohono O’odham Nation.







\appendix

\begin{table*} 
\begin{center}
\begin{threeparttable}[b]
\caption{Modules and their parameters adopted for SED fitting with \texttt{CIGALE}}
    \begin{tabular}{ccc}
    \hline
    \hline
    Modules & Parameters\tnote{a} & Values \\
    \hline 
    \multicolumn{3}{c}{\textbf{Star Formation History}} \\  
    \texttt{sfhdelayed} & \textit{tau\_main}~(Myr) & $100, 250, 500, 750, 1000, 4000, 8000$ \\    
                        & \textit{age\_main}~(Myr) & $1000, 3000, 4500, 6000, 8000,10000,13000$ \\    
                        & \textit{age\_burst}~(Myr) & $50$ \\    
                        & \textit{f\_burst} & $0.0, 0.01$ \\
    \hline
    \multicolumn{3}{c}{\textbf{Stellar Populations}} \\      
    \texttt{bc03} \citep{BC2003} & \textit{imf} & \citet{Chabrier2003} \\
                                 & \textit{metallicity}  & $0.02, 0.05$ \\
    \hline
    \multicolumn{3}{c}{\textbf{Nebular Emission}} \\      
    \texttt{nebular}    & \textit{logU} & $-2.0$ \\
                        & \textit{zgas}  & $0.02$ \\
    \hline
    \multicolumn{3}{c}{\textbf{Attenuation Laws}} \\      
    \texttt{dustatt\_modified\_starburst}   & \textit{E\_BV\_lines} & $0,0.05,0.15,0.3,0.5,0.75,0.9$ \\    
    \citep{Calzetti2000} &  & \\    
                                            & \textit{E\_BV\_factor} & $0.44$ \\    
    \hline
    \multicolumn{3}{c}{\textbf{Dust Emission}} \\      
    \texttt{dl2014} \citep{Draine2014}  & \textit{qpah} & $0.47, 1.12, 2.5,3.19$ \\    
                                        & \textit{umin} & $15.0$ \\       
                                        & \textit{gamma} & $0.02$ \\    
    \hline
    \multicolumn{3}{c}{\textbf{Active Galactic Nuclei}} \\      
    \texttt{fritz2006} \citep{Fritz2006}    & \textit{tau} & $6.0$ \\    
                                            & \textit{gamma} & $0.0$ \\       
                                            & \textit{psy} & $0.001,20.1,40.1,70.1,89.99$ \\       
                                            & \textit{fracAGN} & $0.0,0.01,0.1,0.3,0.5,0.7,0.9$ \\    
    \hline
    \end{tabular}
\begin{tablenotes}
\item [a] Details regarding the parameters can be found in \citet{CIGALE}. We keep the default values for parameters not listed in the table. 
\end{tablenotes}  
\label{cigale_para}
\end{threeparttable}
\end{center}
\end{table*}

\section{Parameter settings for SED fitting} \label{cigale}
We derive stellar masses by performing SED fitting with \texttt{CIGALE} on all the LRGs in this study. 
The adopted modules and parameters are summarized in Table~\ref{cigale_para}. 
Further details on the stellar mass estimation with SED fitting can be found in Section~\ref{stellar}.

\section{LRG Covering Fractions: Color Dependence and Comparison With Previous Studies} 
\subsection{Covering fraction for the blue and red LRGs} 
\label{blue_vs_red}

\begin{figure*} 
\begin{center}
\includegraphics[width=0.9\linewidth]{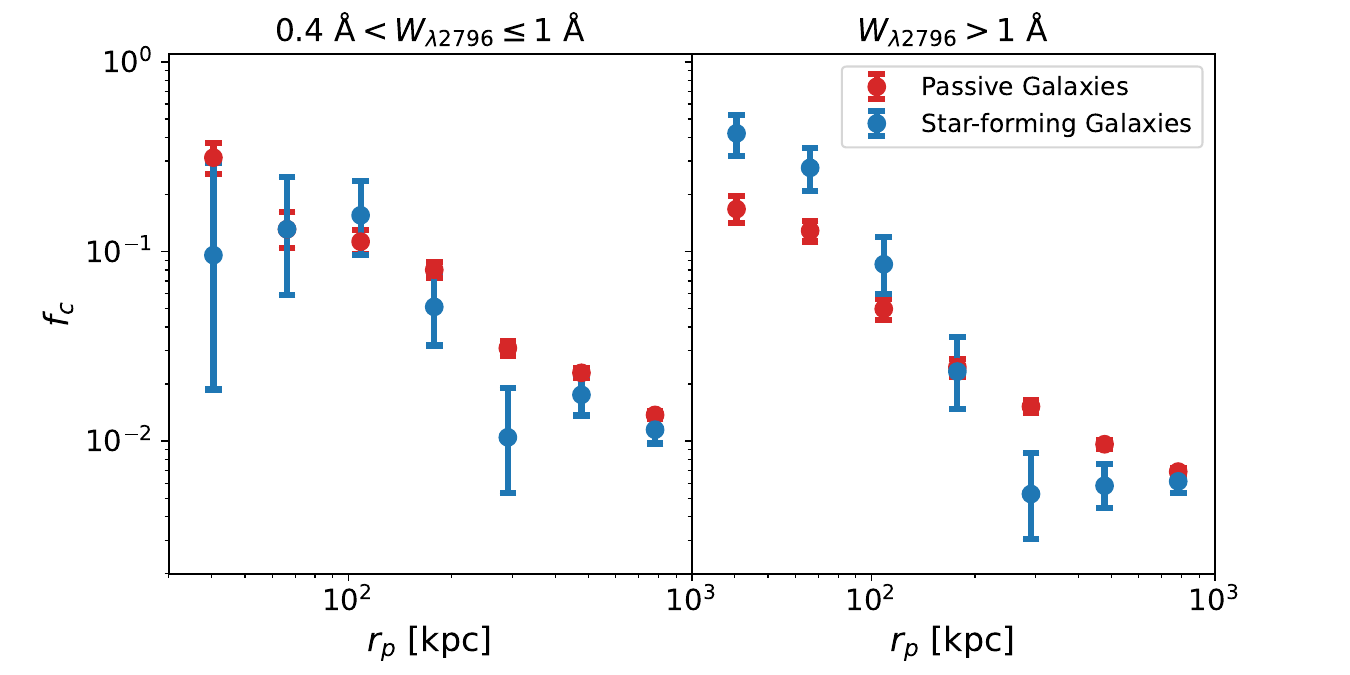}
\end{center}
\caption{Covering fraction ($f_c$) as a function of impact parameter ($r_p$) for SFR-enhanced and quiescent LRGs. The left and right panel shows the $f_c$ for absorbers with $0.4 \leq W_{0,\lambda2796}<1 \rm \, \AA$ (weak absorbers) and that for absorbers with $W_{0,\lambda2796}\geq 1 \rm \, \AA$ (strong absorbers). The errors represent the $1-\sigma$ confidence levels estimated using the binomial statistics \citep{Gehrels1986}.}
\label{cov_rp_blue_red}
\end{figure*}

The covering fraction for both quiescent and SFR-enhanced LRGs as a function of ${\rm r}_p$ in physical space is illustrated in Figure~\ref{cov_rp_blue_red}, separated into strong ($W_{0,\lambda2796} \geq 1 \text{\AA}$) and weak ($0.4 \leq W_{0,\lambda2796} < 1 \text{\AA}$) absorbers. 
Here we select the two LRG populations within the same redshift range $(0.5<z<0.85)$ and stellar mass range $(11.0 < {\rm log_{10}~ M_*/M_\odot} < 11.4)$. 
The $f_c$ of strong absorbers around SFR-enhanced LRGs is higher than that around quiescent LRGs at $r_p \lesssim 100~{\rm kpc}$. 
This agrees with the finding of \citet{Lan14}, who identified a correlation between Mg~{\small II} absorption strength and galaxy star formation activities. 
No significant difference in $f_c$ is observed for weak absorbers between SFR-enhanced and quiescent LRGs. 
As SFR-enhanced LRGs account for only $\sim 9\%$ of the full DESI LRGs, and our interest lies in truly passive galaxies, we discuss only the gas properties of quiescent LRGs in the rest of this paper. 
Hereafter, we refer to ``quiescent LRGs” simply as ``LRGs”. 

\subsection{Comparison with previous studies} \label{compare_previous}

\begin{figure*} 
\begin{center}
\includegraphics[width=0.9\linewidth]{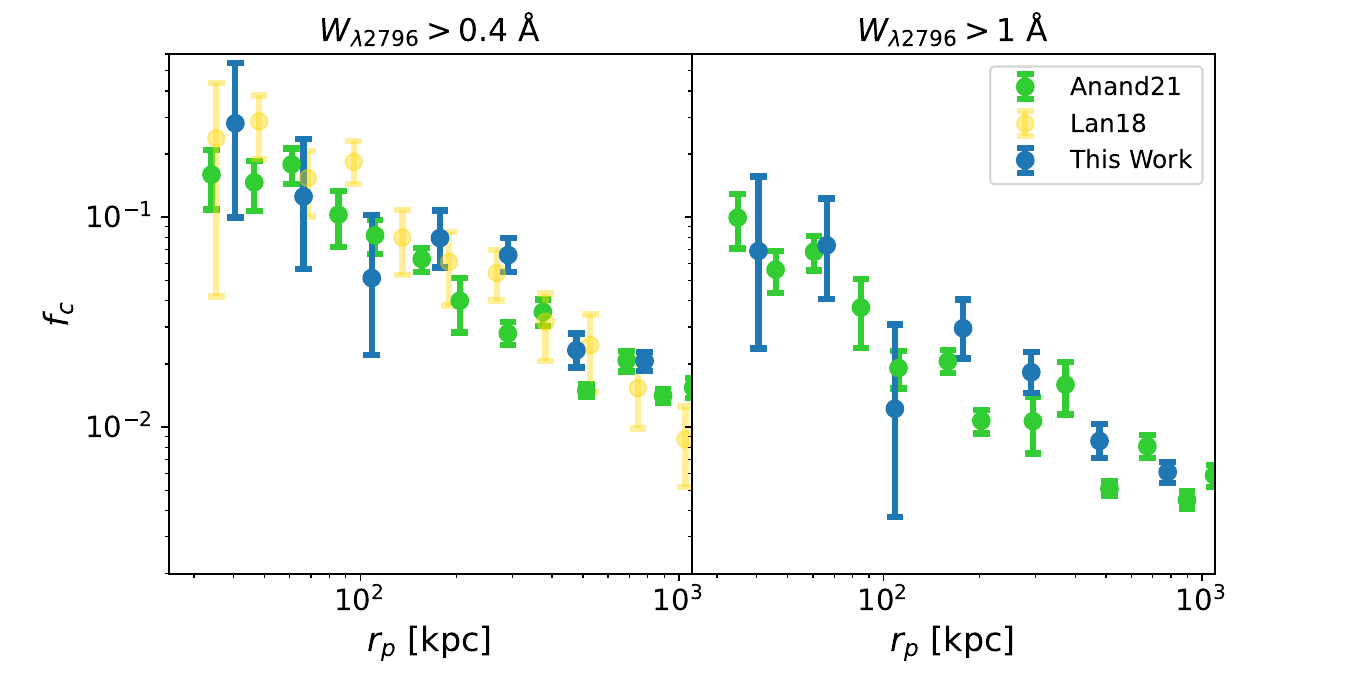}
\end{center}
\caption{Comparison of the DESI covering fraction measured in this work with the SDSS measurement in \citet{Lan2018, Anand2021} for LRGs. Left: $f_c$ in physical scale for absorbers with $W_{0,\lambda2796}\geq 0.4 \rm \, \AA$. Right: $f_c$ for absorbers with $W_{0,\lambda2796}\geq 1 \rm \, \AA$. The blue, green, and yellow circles show the $f_c$ from this work, \citet{Anand2021}, and \citet{Lan2018}, respectively. }
\label{fc_compare}
\end{figure*}

We compare our measurement of $f_c$ ($W_{0,\lambda2796}>0.4 \, \rm \AA$ and $W_{0,\lambda2796}>1 \, \rm \AA$) of DESI LRGs with the measurements of SDSS LRGs in Figure~\ref{fc_compare}. 
We note that the DESI LRGs exhibit a higher median redshift and a smaller median stellar mass than the SDSS LRGs \citep{Chen2012, Lan2018, SDSSDR14, SDSSDR16}. 
To perform a fair comparison, we select a subsample of $\sim 40,000$ DESI LRGs whose median redshift and stellar mass are comparable to those of the SDSS LRGs. 
Specifically, the subsample has a median redshift of $z\approx 0.55$ and a median stellar mass of $\log_{10}\,{\rm M_{halo}}/{\rm M_\odot} = 11.3$. 
Background absorber contributions have been subtracted from the covering fractions in both our analysis and that of \citet{Anand2021}. 

Figure~\ref{fc_compare} shows that the $f_c$ of the DESI LRGs is consistent with that of the SDSS LRGs from \citet{Lan2018} and \citet{Anand2021} at a given impact parameter, provided that the three samples have similar redshift and stellar mass distributions. 
All of them show comparable decreasing trends with similar slopes. 
We emphasize that the three $f_c$ measurements presented here are derived using distinct methods and catalogs, including direct fitting of individual quasar spectra with galaxy redshifts incorporated (Our work), stacking of background quasar spectra before fitting \citep{Lan2018}, and matching of detected absorbers with LRG catalogs \citep{Anand2021}.


Our finding of a redshift trend in $f_c$ is consistent with that of \citet{Lan2020} for absorbers with $W_{0,\lambda2796} \geq 1\text{\AA}$. 
They reported a redshift dependence $(1+z)^{\alpha}$ with $\alpha=4.0\pm0.4$ for strong absorbers within $r_{vir}$, while weak absorbers ($0.4 \leq W_{0,\lambda2796} < 1,\text{\AA}$) with $\alpha = 0.8 \pm 0.4$. 
The differences may be due to the fact that the passive galaxies in \citet{Lan2020} have a smaller median stellar mass. 
They included passive galaxies with stellar masses as low as ${\rm log}_{10}({\rm M_*}/{\rm M_\odot})=10-10.5$, while only a few sources in our sample fall into this range. 
As shown in Figure~\ref{cov_para_norm}, the redshift dependence is weaker in less massive galaxies, and the generally lower SNR of weak absorbers further hinders the detection of this trend.

\section{Number of LRG-QSO pairs} \label{pair_number}
Table~\ref{cov_pair_number_physical} lists the number of LRG--quasar pairs contributing to each $r_p$ bin for different stellar-mass and redshift bins, in both physical and halo-scaled units.

\begin{table*} 
\begin{center}
\caption{Number of pairs for covering fraction studies. Values without parentheses correspond to physical $r_p$ bins, while values in parentheses correspond to halo-scaled $r_p/r_{\rm vir}$ bins.}
    \begin{tabular}{ccccccc}
    \hline
      & \multicolumn{3}{c}{$0.4 \leq W_{0,\lambda2796} < 1~{\rm \AA}$} & \multicolumn{3}{c}{$W_{0,\lambda2796} \geq 1~{\rm \AA}$} \\
     $\log_{10}\,{\rm M}_{*}/{\rm M}_{\odot}$ & $10.2 - 11.1$& $11.1-11.3$& $11.3-12.3$& $10.2 - 11.1$& $11.1-11.3$& $11.3-12.3$ \\
    \hline   
    \hline
    \multicolumn{7}{c}{$0.4\leq {\rm z} < 0.6$} \\
    $32-52\,{\rm kpc} (0.05-0.09\,r_{rvir})$ &7(2)  &23(17)  &11(14)  &19(4) &59(34)  &35(38) \\
    $52-85\,{\rm kpc} (0.09-0.15\,r_{rvir})$ &16 (3)  &43(19)  &29(36)  &33(14)  &102(66)  &85(101) \\
    $85-139\,{\rm kpc} (0.15-0.27\,r_{rvir})$  &28 (20)  &87 (68)  &71 (99)  &69 (42)  &248 (176)  &191 (295) \\
    $139-228\,{\rm kpc} (0.27-0.47\,r_{rvir})$   &74 (42)  &233 (217) &195 (402)  &208 (105)  &709 (600)  &559 (1125) \\
    $228-373\,{\rm kpc} (0.47-0.82\,r_{rvir})$  &248 (129)  &806 (746)  &685 (1390)  &676 (361) &2238 (2142)  &1942 (3986) \\
    $373-611\,{\rm kpc} (0.82-1.43\,r_{rvir})$  &709 (487)  &2409 (2718)  &1962 (5028)  &1944 (1289)  &6677 (7495)  &5584 (14253) \\
    $611-1000\,{\rm kpc} (1.43-2.5\,r_{rvir})$  &2007(1535)  &7049(900)  &6173(16197)  &5749(4333) &20031 (25630)  &17420 (45677) \\
     \hline
    \multicolumn{7}{c}{$0.6\leq {\rm z} < 0.8$} \\
    $32-52\,{\rm kpc} (0.05-0.09\,r_{rvir})$ & 12 (0)   &23 (13)  &14 (14) &33 (5)  &62 (26)  &39 (38) \\
    $52-85\,{\rm kpc} (0.09-0.15\,r_{rvir})$   &17 (10)  &38 (14)  &24 (27)  &47 (23) &123 (54) &75 (73) \\
    $85-139\,{\rm kpc} (0.15-0.27\,r_{rvir})$  &50 (18)  &125 (76) &74 (86) &125 (51) &342 (216) &195 (240) \\
    $139-228\,{\rm kpc} (0.27-0.47\,r_{rvir})$  &131 (59)  &387 (255) &198 (254) &360 (146) &969 (630)  &530 (692) \\
    $228-373\,{\rm kpc} (0.47-0.82\,r_{rvir})$  &392 (174) &1113 (782) &642 (999) &1091 (488) &2972 (2104)  &1690 (2654) \\
    $373-611\,{\rm kpc} (0.82-1.43\,r_{rvir})$  &1279 (656) &3390 (3006) &1889 (3453) &3363 (1734) &9065 (7989)  &4938 (9078) \\
    $611-1000\,{\rm kpc} (1.43-2.5\,r_{rvir})$  &3763 (2243) &10447 (10351) &5800 (11770)  &10184 (6038)  &28026 (27831)  &15258 (31216) \\
    \hline
    \multicolumn{7}{c}{$0.8\leq {\rm z} < 0.9$} \\    
    $32-52\,{\rm kpc} (0.05-0.09\,r_{rvir})$ &  11 (4)  &22 (11)  &16 (9)  &44 (11)  &73 (27)  &42 (23) \\
    $52-85\,{\rm kpc} (0.09-0.15\,r_{rvir})$   &29 (1)  &65 (12)  &28 (28)  &89 (11)  &143 (48)  &71 (67) \\
    $85-139\,{\rm kpc} (0.15-0.27\,r_{rvir})$ &84 (24)  &147 (85)  &66 (54)  &244 (80)  &375 (198)  &214 (185) \\
    $139-228\,{\rm kpc} (0.27-0.47\,r_{rvir})$  &277 (92)  &436 (246)  &231 (248)  &710 (269)  &1096 (616)  &578 (598) \\
    $228-373\,{\rm kpc} (0.47-0.82\,r_{rvir})$  &830 (321)  &1333 (775)  &700 (819)  &2159 (835)  &3443 (2026)  &1874 (2210) \\
    $373-611\,{\rm kpc} (0.82-1.43\,r_{rvir})$ &2443 (1138)  &4020 (2860)  &2148 (2967)  &6364 (2968)  &10624 (7501)  &5756 (7910) \\
    $611-1000\,{\rm kpc} (1.43-2.5\,r_{rvir})$  &7505 (3899)  &12079 (9867)  &6509 (10152)  &19774 (10217)  &32140 (26190)  &17318 (27159) \\     
    \hline 
    \multicolumn{7}{c}{$0.9\leq {\rm z} < 1.2$} \\    
    $32-52\,{\rm kpc} (0.05-0.09\,r_{rvir})$ &5 (0)  &12 (4)  &18 (8)  &16 (1)  &30 (11)  &42 (17) \\
    $52-85\,{\rm kpc} (0.09-0.15\,r_{rvir})$   &15 (2)  &25 (6)  &35 (21)  &41 (9)  &70 (16)  &83 (50)   \\
    $85-139\,{\rm kpc} (0.15-0.27\,r_{rvir})$ &44 (10)  &54 (29)  &61 (59)  &115 (22)  &189 (80)  &160 (142) \\
    $139-228\,{\rm kpc} (0.27-0.47\,r_{rvir})$  &134 (33)  &186 (71)  &181 (132)  &360 (91)  &500 (235)  &490 (371) \\
    $228-373\,{\rm kpc} (0.47-0.82\,r_{rvir})$  &399 (125)  &539 (272)  &581 (542)  &1092 (332)  &1529 (762)  &1577 (1403) \\
    $373-611\,{\rm kpc} (0.82-1.43\,r_{rvir})$  &1374 (429)  &1870 (1065)  &1731 (1853)  &3655 (1154)  &5108 (2907)  &4801 (5116) \\
    $611-1000\,{\rm kpc} (1.43-2.5\,r_{rvir})$  &4241 (1689)  &6152 (4083)  &5583 (6772)  &11433 (4542)  &16710 (11042)  &15373 (18578) \\
    \hline
    \end{tabular}
\label{cov_pair_number_physical}
\end{center}
\end{table*}

\section{Contribution from random Mg~{\small II} absorbers in the background} \label{background}
To account for the contribution from background Mg~{\small II} absorbers, we build a random LRG sample and estimate the expected mean absorption signal (e.g., $f_{c,b}$) along random sightlines. 
The position and redshift of the random sample are shuffled from the DESI year 1 LRG sample, keeping the same number of sources as the observed sample but breaking the relation between the position and redshift. 
The redshift and position distributions of the random LRG sample are similar to those of the observed LRGs. 
The background quasar sample for the random test remains the same. 
We pair and select the random LRGs with background quasars as in Section~\ref{build_sample}, leading to a number of $70,058$ random pairs, with a median redshift of $0.744$ for the random LRGs. 
Among these pairs, $3104$ background absorbers were detected. 

\begin{figure*} 
\begin{center}
\includegraphics[width=1.\linewidth]{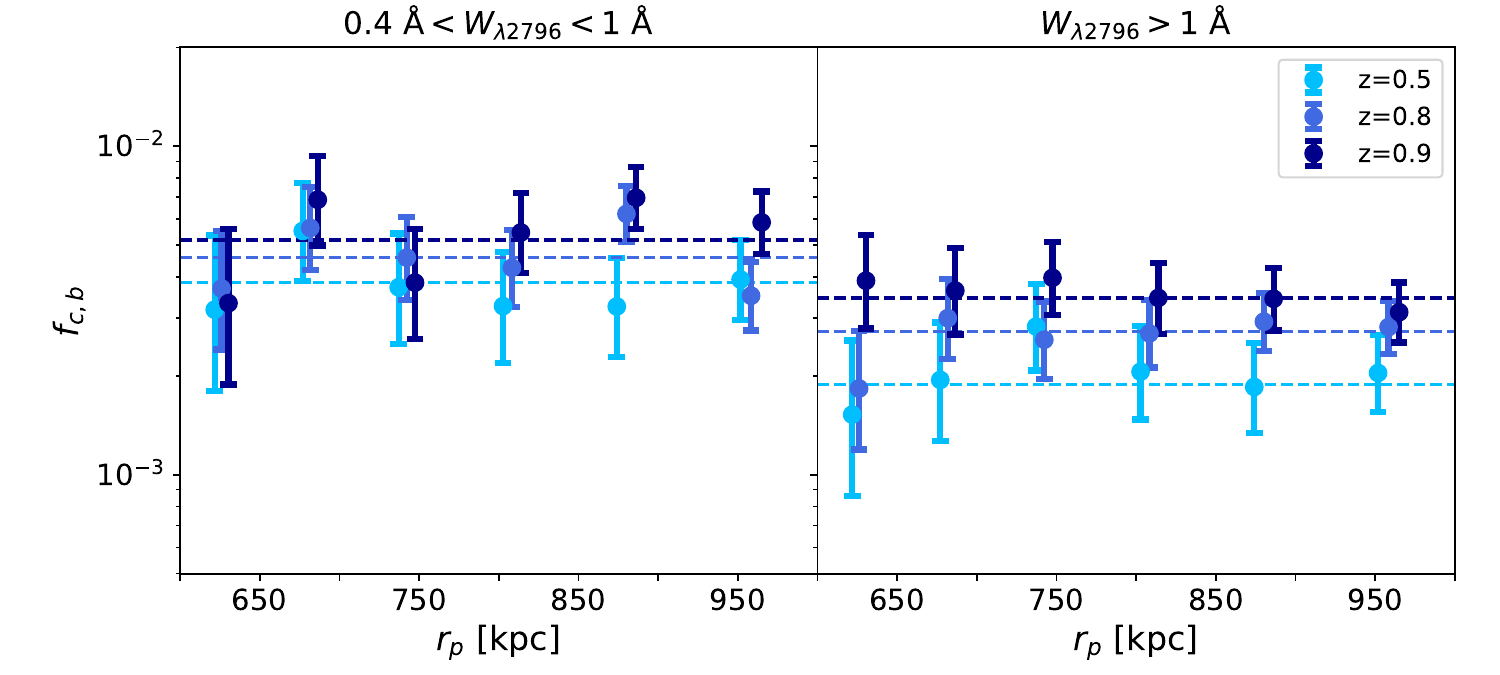}
\end{center}
\caption{Random covering fraction as a function $r_p$ in physical space for weak (Left) and strong (Right) absorbers. Colors indicate the redshift of galaxies. The dashed lines are the expected values from \citet{Zhu2013b}.}
\label{random_fc}
\end{figure*}

The random covering fraction varies with redshift \citep{Zhu2013b}, and thus our random $f_{c,b}$ was estimated using the same redshift bins as the observed $f_{c,obs}$. 
To avoid incompleteness, we consider only pairs with background quasar spectra having $SNR \geq 8$ for the random sample. 
Figure~\ref{random_fc} shows the random covering fraction in three redshift ranges.
The covering fraction from background sources is independent of the impact parameter, and our result is consistent with that in \citet{Zhu2013b}. 
We use the median covering fraction across $r_p$ as the random contribution subtracted from the observed $f_{c,obs}$.

\begin{figure} [h!]
\begin{center}
\includegraphics[width=0.7\linewidth]{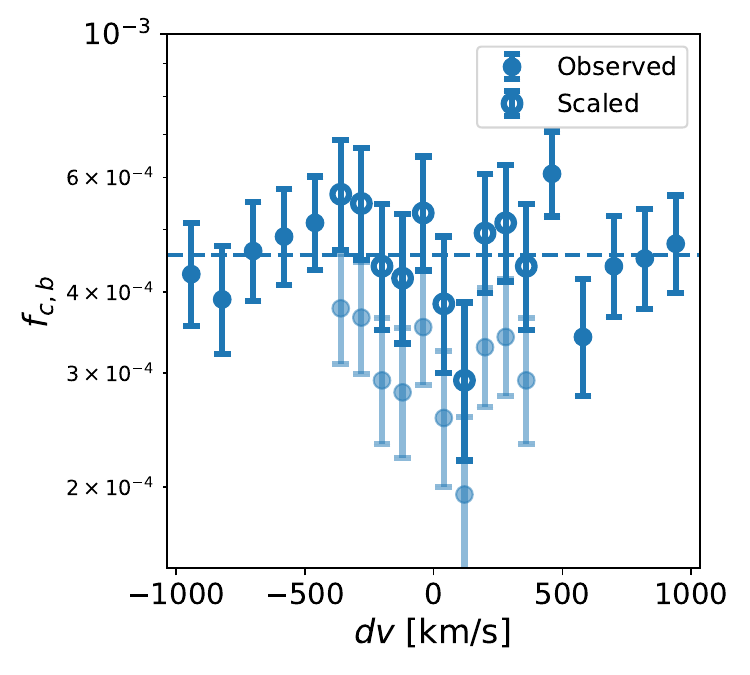}
\end{center}
\caption{Random covering fraction for absorbers with $W_{0,\lambda2796} \geq 1~{\rm \AA}$ in velocity space. Filled circles show the observed $f_{c,b}$, and open circles show $f_{c,b}$ scaled by the bin size ratio. Semi-transparent data points represent the original values before bin-size scaling. The dashed line indicates the median random $f_{c,b}$ for absorbers with $|dv|>400~{\rm km,s^{-1}}$.}
\label{random_fc_vspace}
\end{figure}

We next consider the background contribution to the LoS velocity distribution. 
To do so, we estimate the random $f_{c,b}$ for absorbers in the velocity space, as shown in Figure~\ref{random_fc_vspace}. 
That is, the covering fraction measured at a fixed impact parameter for different velocity ranges. 
To better resolve the dispersion at lower velocities, we adopt variable bin sizes: $80~{\rm km/s}$ for absorbers with $|dv|< 400~{\rm km/s}$ and $120~{\rm km/s}$ for absorbers with $|dv|> 400~{\rm km/s}$. 
We weight the $f_{c,b}$ values in the lower velocity bins by $120/80$ in Figure~\ref{random_fc_vspace} to account for this difference, yielding a uniform distribution of $f_{c,b}$ across velocity ranges after scaling. 
As was done in the analysis of gas dispersion with the observed sample, we did not divide the galaxies into several redshift bins. 
The random background contribution is estimated separately for the low- ($|dv|<400~{\rm km,s^{-1}}$) and high-velocity ($|dv|>400~{\rm km,s^{-1}}$) ranges using their respective median $f_{c,b}$ values.  

\section{Gas velocity as a function of redshift} \label{veldisp_zdep}
We apply the KS test to evaluate whether the gas velocity distributions differ across redshift for each bin, as illustrated in Figure~\ref{vel_dist_hist}. 
Table~\ref{vdist_pval} summarizes the p-values of the KS test, suggesting that the gas velocity does not depend on the redshift of absorbers. 
Although the KS test yields a low p-value ($\sim 0.004$) for the bin $10.2 \leq \log_{10}\,{\rm M_*}/{\rm M_\odot} < 11.1$, suggesting a potential difference in velocity distributions between high-redshift and low-redshift absorbers, the gas velocity dispersion (Figure~\ref{mad_zdep}) appears consistent across redshifts. 
The small p-value observed in that bin appears to be driven by differences in the distribution of absorbers at large velocity offsets.
When applying a velocity cut of $|dv|\leq 500~{\rm km/s}$, the resulting p-value increases to $\sim 0.1$, consistent with the absence of a redshift dependence. 
In conclusion, our results show no relationship between the gas dispersion and redshift. 

\begin{table} [h!]
\caption{P-value of the KS test for velocity distribution.}
\begin{center}
    \begin{tabular}{c|ccc}
    \hline
    \hline
     & $r_p \leq 0.5 r_{vir}$ & $0.5 r_{vir} < r_p \leq r_{vir}$ & $r_{vir} < r_p \leq 2.5 r_{vir}$ \\
    \hline
    $10.2 < \log_{10}\,{\rm M}_{*}/{\rm M}_{\odot} \leq 11.1$ & $0.43$ & $0.14$ & $0.004$ \\
    $11.1  < \log_{10}\,{\rm M}_{*}/{\rm M}_{\odot} \leq 11.3$ & $0.94$ & $0.26$ & $0.29$\\
    $11.3  < \log_{10}\,{\rm M}_{*}/{\rm M}_{\odot} \leq 12.3$ & $0.62$ & $0.85$ & $0.42$\\
    \hline
    \end{tabular}
\label{vdist_pval}
\end{center}
\end{table}

\section{Fitting the velocity dispersion} \label{fit_velocity}
\begin{figure*} 
\begin{center}
\includegraphics[width=1.\linewidth]{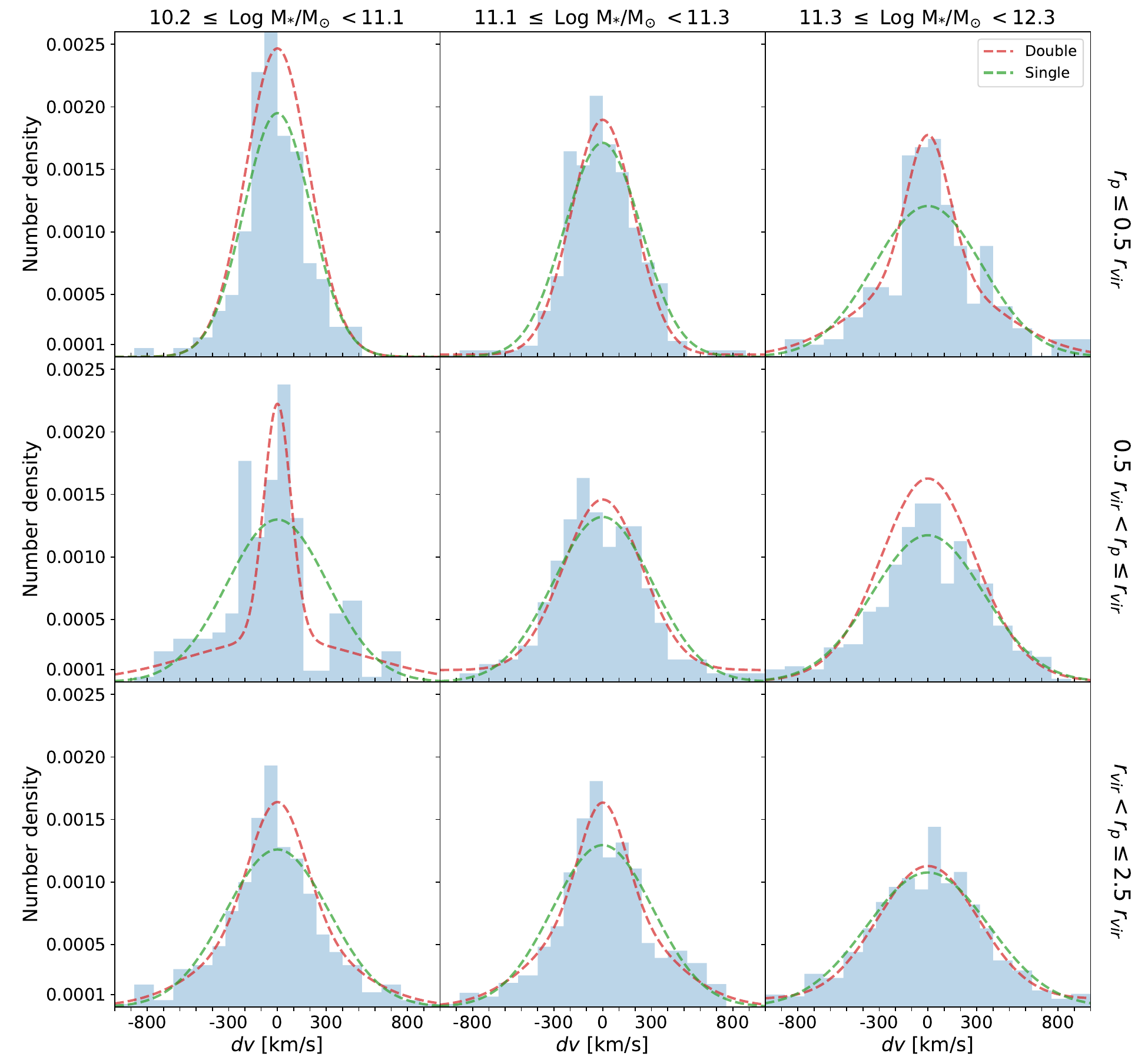}
\end{center}
\caption{Comparison between the fits with single Gaussian and double Gaussian. The blue shaded area shows the velocity distributions of LRG Mg~{\small II} absorbers in each stellar mass and impact parameter bin. Green and red curves correspond to the posterior-median single- and double-Gaussian models.}
\label{vel_disp_model_compare}
\end{figure*}

We estimate the gas dispersion by fitting the velocity distributions, as described in Section~\ref{vel_disp}. 
Figure~\ref{vel_disp_model_compare} presents single- and double-Gaussian fits, illustrating that the double-Gaussian model more effectively captures the velocity distribution in each bin. 
We quantify this using the BIC, showing that the double-Gaussian model provides a better statistical description with $\Delta{\rm BIC}={\rm BIC~(single-Gaussian)}-{\rm BIC~(double-Gaussian)} \gtrsim 6$.

\begin{figure*} 
\begin{center}
\includegraphics[width=1.\linewidth]{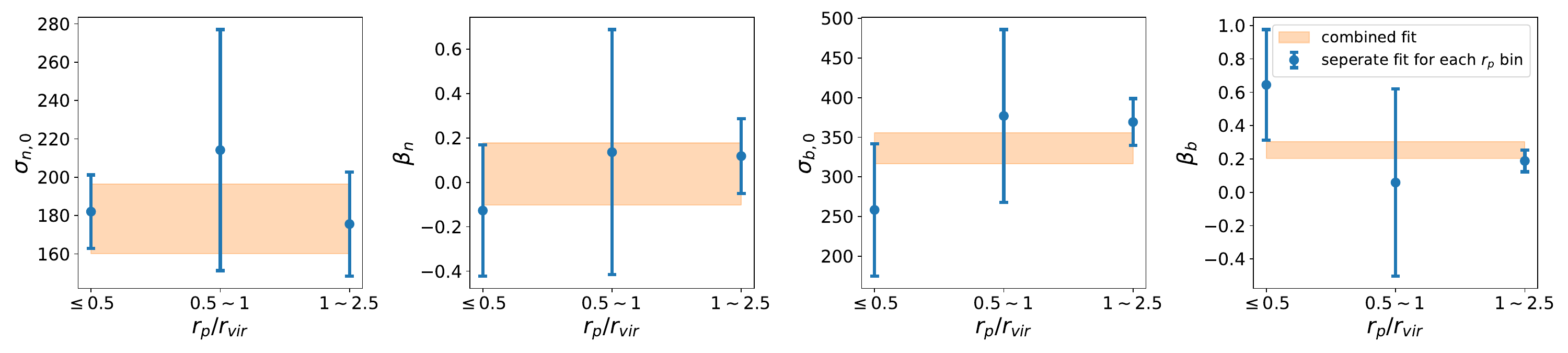}
\end{center}
\caption{The Posterior median and 68\% credible interval of the global fit parameters for velocity distribution. The blue circles represent the results from individual fits in each $r_p$ bin, while the orange shaded regions show the results from the combined $r_p$ fit.}
\label{vel_disp_global_para}
\end{figure*}

To better constrain the velocity dispersion, we perform a global fit across stellar mass bins, with additional parameters, $\beta$, in Equation~\ref{fc_vel_gauss} introduced to capture the dependence of $\sigma$ on stellar mass. 
Given that the dispersion may also depend on impact parameter, we fit the velocity distribution separately for each $r_p$ bin to examine potential differences in the regression parameters.
Figure~\ref{vel_disp_global_para} shows the fitted parameters as a function of impact parameter.

As shown in the figure, neither $\sigma$ nor $\beta$ exhibits a significant dependence on impact parameter.
Accordingly, a combined global fit across all impact parameters is performed to improve parameter constraints. 
Figure~\ref{vel_disp_global_para} shows that the combined fit results for $\sigma$ and $\beta$ are consistent with those obtained from individual fits to each $r_p$ bin.

\bibliography{sample631}{}
\bibliographystyle{aasjournal}



\end{document}